%% file: acl_latex.tex
\documentclass[11pt]{article}

\usepackage[final]{acl}

\newcommand{\wmkey}{\mathsf{K}}

\usepackage{times}
\usepackage{latexsym}
\usepackage[T1]{fontenc}
\usepackage[utf8]{inputenc}
\usepackage{microtype}
\usepackage{inconsolata}
\usepackage{graphicx}
\usepackage{amsmath}
\usepackage{booktabs}      
\usepackage{colortbl}      
\usepackage{multirow}      
\usepackage{array}         
\usepackage{amsmath}       
\usepackage{amssymb}       
\usepackage{amsthm}        
\usepackage[dvipsnames]{xcolor}  
\usepackage{enumitem}      
\usepackage{algorithm}     
\usepackage{algpseudocode} 
\usepackage{placeins}      
\usepackage{tikz}          
\usetikzlibrary{arrows.meta,positioning,shapes,fit,calc}
\usepackage[most]{tcolorbox}  
\usepackage{pifont}        
\usepackage{makecell}      
\usepackage{siunitx}       
\theoremstyle{plain}
\newtheorem{lemma}{Lemma}

\theoremstyle{definition}

\newcommand{\nonceT}{\mathit{nonce}_t}
\newcommand{\carrier}{\mathrm{carrier}}
\newcommand{\candset}{\mathcal{C}}
\newcommand{\probdist}{\pi}
\newcommand{\ctx}{\mathrm{ctx}}
\newcommand{\commit}{\mathrm{cm}}
\newcommand{\header}{\mathrm{hdr}}

\usepackage[table]{xcolor}
\definecolor{cbblue}{RGB}{198,219,239}

\newcommand{\amem}{\textsc{A-Mem}}
\newcommand{\graphiti}{\textsc{Graphiti}}
\newcommand{\memmark}{\textsc{MemMark}}
\newcommand{\agentmark}{\textsc{AgentMark}}
\newcommand{\kgmark}{\textsc{KGMark}}
\newcommand{\tiermem}{\textsc{TierMem}}
\newcommand{\locomo}{LoCoMo}

\title{\memmark: State-Evolution Attribution Watermarking for Agent Long-Term Memory Systems}

\author{
  \normalfont\normalsize
  \begin{tabular}{c}
    \textbf{Haobo Zhang$^{1,*}$ \quad Xutao Mao$^{2,*}$} \\
    \textbf{Guangyuan Dong$^{3}$ \quad Ziwei Li$^{4}$ \quad Xuanbo Su$^{5}$} \\
    \textbf{Kaijie Chen$^{6}$ \quad Jing Yang$^{7}$ \quad Zheng Lin$^{8,\dagger}$} \\
    $^{1}$Zhejiang University of Technology, $^{2}$Independent Researcher \\
    $^{3}$National University of Singapore, $^{4}$King Abdullah University of Science and Technology \\
    $^{5}$Bairong, $^{6}$Tongji University, $^{7}$Universiti Malaya, $^{8}$University of Hong Kong \\
    \texttt{zhanghaobo@zjut.edu.cn} \\
    \textsuperscript{*}Equal contribution. \quad
    $^{\dagger}$Corresponding author.
  \end{tabular}
}

\begin{document}
\maketitle

\input{sections/abstract}
\input{sections/introduction}
\input{sections/related_work}

\input{sections/method}
\input{sections/experiments}
\input{sections/conclusion}
\input{sections/limitation}

\bibliography{main}

\appendix
\input{sections/appendix}

\end{document}

%% file: sections/abstract.tex
\begin{abstract}
Memory-backed agents need provenance that can survive leaked or migrated
snapshots, where logs, visible outputs, and trusted metadata may be absent.
We propose \memmark{}\footnote{\url{https://github.com/zhb0119/MemMark}}, a state-evolution attribution watermark that embeds an
owner-controlled signal into latent memory-write decisions. At each internal
LLM call, \memmark{} samples among admissible candidates using keyed,
distribution-preserving selection, and records cryptographic commitments with
signed session anchors and reveal evidence. This makes attribution depend on
reproducible backend behavior rather than mutable provenance fields. Across \amem{} and \graphiti{} on \locomo{}, with three LLM backbones,
\memmark{} preserves memory utility: Overall F1 retains 99.6\% of the
unwatermarked baseline, while BLEU-1 changes by +0.2\%. It also provides
usable carrier capacity, with 1.16, 1.14, and 1.26 bits of mean entropy for
update-target, link-target, and semantic-realization decisions. In the
snapshot-only R3 setting, \memmark{} recovers the full 40-bit payload from
final snapshots, while wrong-key verification remains near chance. Under nine
memory-lifecycle attacks, verification distinguishes tampering, evidence
deletion, and partial payload recovery. These results show that robust
snapshot-only attribution is feasible for long-term agent memory without
surviving traces, trusted metadata, or utility-degrading.
\end{abstract}

%% file: sections/introduction.tex
\section{Introduction}
\label{sec:intro}

LLM agents are increasingly moving from single-session responders to
persistent actors whose decisions depend on state that survives across
interactions. In such systems, the memory layer is no longer a passive
cache: it becomes part of the security boundary~\cite{park2023generative,
memgpt,memorybank,chen2024agentpoison,amemguard,mnemonicsovereignty}.
Recent systems such as \amem{}, \graphiti{}, Mem0, MemOS, Memory-R1, and
MemMachine~\cite{amem,graphiti,chhikara2025mem0,memos,yan2025memory,memmachine} maintain
long-lived state through extraction, updates, consolidation, linking, and deletion.
Benchmarks likewise evaluate memory through long-horizon recall,
knowledge updates, temporal reasoning, and incremental multi-turn state
maintenance~\cite{maharana-etal-2024-evaluating,longmemeval,
memoryagentbench}. As memory writing becomes an explicit object of system
design~\cite{mem-as-action,a-mac,prefmem}, a natural response is to attach
provenance fields, such as source anchors, versions, or lifecycle
traces~\cite{tiermem,memos}. These fields are useful when the writer and
storage layer are trusted, but they are much less useful when the memory
snapshot itself may have been rewritten.

We study a post-compromise forensic setting in which the verifier may not
have access to a trusted write-time trace. Prompts, tool calls,
memory-write requests, and backend logs may be absent, incomplete, or
controlled by the same actor who controls the memory store; the only
durable artifact may be the final memory snapshot. This setting is not a
replacement for trusted logging in normal operation, but the fallback case
that remains when such logs are lost, withheld, or suspected to be
corrupted. An attacker can rewrite ownership fields, erase identifiers,
fabricate provenance chains, or edit backend-native histories such as
\amem{} evolution logs and \graphiti{} fact-invalidation traces. This
threat model is motivated by evidence that agent memory can be poisoned,
stealthily modified, or persistently compromised~\cite{chen2024agentpoison,
amemguard,memorygraft,mnemonicsovereignty}, and by recent surveys that identify
agent memory attacks as a distinct safety risk for large-model-powered
agents~\cite{ma2026safetyatscale}. Thus, field-based provenance
has a circular failure mode: the same untrusted snapshot contains both the
contested memory and the mutable fields that certify it.

Watermarking offers a natural way to make attribution survive untrusted
handling. Token-level LLM watermarks~\cite{kgw,synthid,markllm,
mao-etal-2025-watermarking}, RAG or structured-data
watermarks~\cite{ragwm,ragdi,canaries,aqua,graphdbwm,kgmark}, and
agent-level behavioral watermarks~\cite{agentguide,agentmark,acthook,
agentwm} provide important attribution mechanisms. However, their signals
are placed in generated text, protected corpora or graphs, visible tool
use, or action trajectories. Long-term memory forensics exposes a
different evidence channel: the original prompts, tool calls, and
execution trace may be unavailable, while the surviving artifact is only
the final memory snapshot. The question is therefore whether memory
evolution itself can leave reproducible evidence of who wrote it.

We propose \memmark{}, a watermark for long-term memory evolution under
adversarial snapshot control. Instead of storing attribution in mutable
metadata, \memmark{} embeds it into latent but utility-preserving
state-transition choices: which existing item to update, which related
item to link, or which semantically equivalent realization to store. A
keyed sampler selects among admissible candidates while preserving the
backend's preference distribution, so attribution is carried by the
backend's own evolution behavior rather than by a self-reported field. To
make this evidence usable after the external execution trace is lost,
\memmark{} records an in-snapshot audit path whose validity is checked by
replay against the keyed sampler and the admissible transition set.
\memmark{} supports three verification regimes: R1 with a complete
external log, R2 with a partial log, and R3 with only the final memory
snapshot. In this way, \memmark{} moves provenance from editable claims to
a reproducible behavioral trace.
\paragraph{Contributions.}
\begin{enumerate}[topsep=2pt,itemsep=2pt,leftmargin=*]
\item \textbf{Snapshot-only attribution with full payload recovery.}
We introduce an audit design and R1/R2/R3 verification hierarchy that supports
attribution even when only the final memory snapshot is available. In R3,
\memmark{} recovers the full 40-bit payload, compared with no recovery for
signed-metadata-only and 15\% wrong-key recovery.

\item \textbf{Utility-preserving memory-evolution watermarking.}
We introduce \memmark{}, a backend-invariant watermarking abstraction that
embeds attribution into latent memory-evolution decisions rather than editable
provenance fields. Across six model--backend settings, \memmark{} retains
99.6\% of the unwatermarked Overall F1 and improved by 0.2\% of BLEU.

\item \textbf{Cross-backend capacity and attack diagnostics.}
Across \amem{} and \graphiti{}, \memmark{} exposes mean carrier entropies of
1.16, 1.14, and 1.26 bits for update-target, link-target, and
semantic-realization decisions, and remains diagnostic under nine
memory-lifecycle attacks at strengths 0.1, 0.3, and 0.5.
\end{enumerate}
\begin{figure*}[t]
\centering
\includegraphics[width=\textwidth]{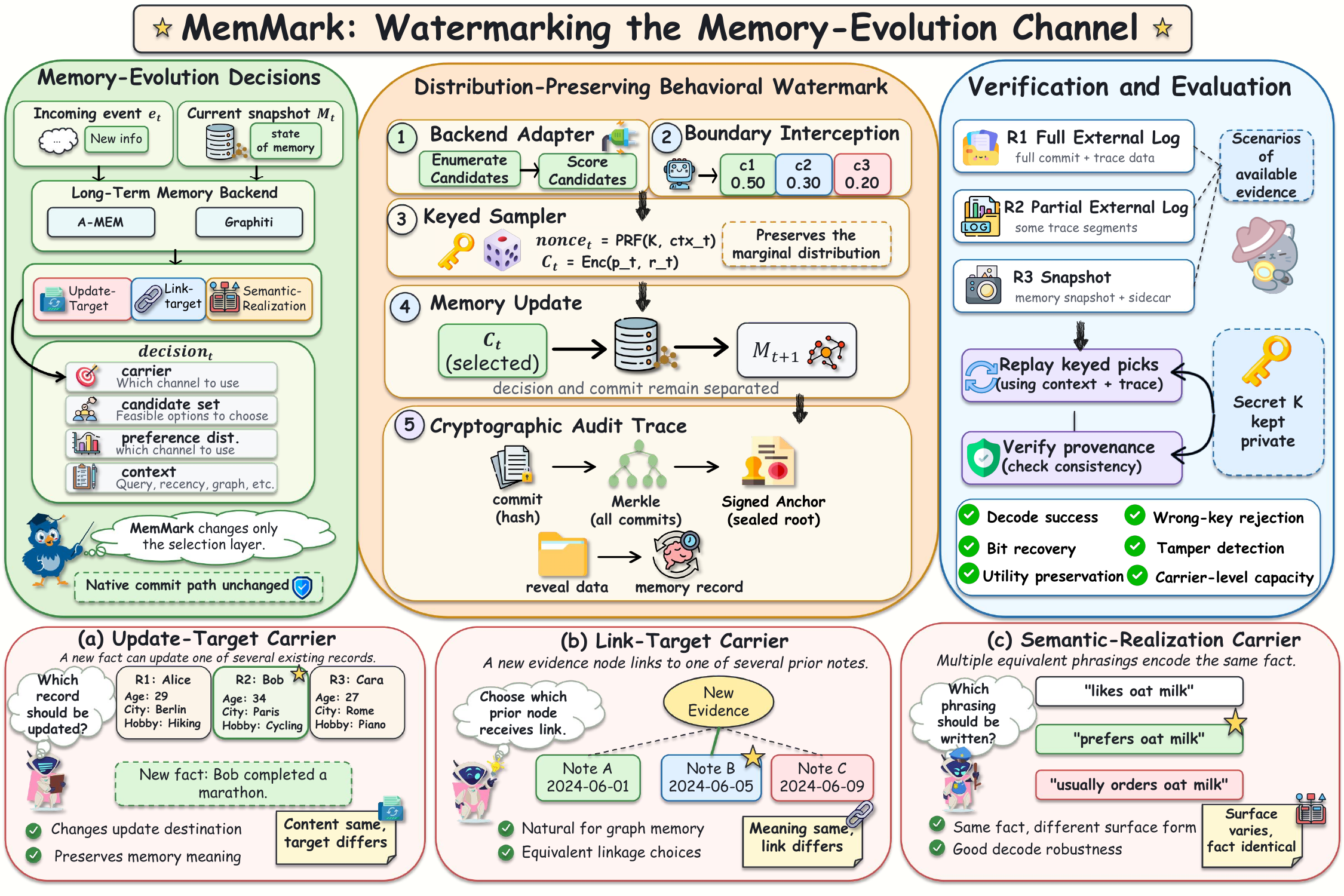}
\caption{\textbf{End-to-end \memmark{} pipeline.} A memory write exposes
carrier-specific choices over update targets, link targets, and semantic
realizations. \memmark{} enumerates and scores feasible candidates,
intercepts the LLM-call boundary, and uses a secret-keyed,
distribution-preserving sampler to select $c^*$ while leaving the native
commit path unchanged. Each selected decision is bound to a commitment,
per-session Merkle log, and signed anchor; reveal data is stored with the
memory record. The same evidence supports R1 full-log, R2 partial-log,
and R3 snapshot-only attribution by replaying keyed picks and checking
provenance consistency.}
\label{fig:pipeline}
\end{figure*}

%% file: sections/related_work.tex
\section{Related Work}
\label{sec:related}

\subsection{Watermarking for Text, Data, and Agent Behavior}
\label{sec:related-watermarking}
Existing watermarks protect evidence channels that may be absent in
memory forensics. Text watermarks embed signals in generated
tokens~\cite{kgw,synthid,markllm,mao-etal-2025-watermarking,
gao2026pathmark,xu2026forgetmark};
data-level methods protect retrieval corpora, multimodal stores, or
graphs~\cite{ragwm,ragdi,canaries,aqua,graphdbwm,kgmark}; and agent
watermarks target visible planning, tool use, or trajectory
data~\cite{agentguide,agentmark,acthook,agentwm}. \memmark{} keeps the
behavioral intuition but moves the carrier to memory-evolution decisions
that remain recoverable from the backend snapshot itself.

\subsection{Long-Term Agent Memory and Adversarial Provenance}
\label{sec:related-memory}
Persistent-memory agents and backends such as \amem{}, \graphiti{},
MemOS, Memory-R1, and MemMachine maintain note networks, temporal graphs,
or structured memory objects~\cite{park2023generative,memgpt,memorybank,
amem,graphiti,memos,yan2025memory,memmachine}; benchmarks evaluate their
long-horizon recall, updates, temporal reasoning, and incremental
multi-turn memory~\cite{maharana-etal-2024-evaluating,longmemeval,
memoryagentbench}. Work on admission, update, and curation treats memory
writing as an explicit decision layer~\cite{mem-as-action,a-mac,prefmem},
which is exactly the layer \memmark{} watermarks.

Provenance and security work motivate the adversarial setting.
\tiermem{} and MemOS attach source anchors, versioning, or lifecycle
metadata~\cite{tiermem,memos}, while poisoning and memory-security
studies show that memory can be maliciously or persistently
compromised~\cite{chen2024agentpoison,amemguard,memorygraft,
mnemonicsovereignty,gao2026bit,yu2025black,
zhang2026attndiffattentionbaseddifferentialfingerprinting,
huang-etal-2026-toposhield}. Metadata helps under trusted storage, but not when
the attacker controls the snapshot containing both memory and claimed
provenance. \memmark{} instead combines replayable keyed choices with
commitments, Merkle trees, and transparency logs~\cite{pedersen1991non,
merkle1987digital,certificate-transparency}.

%% file: sections/method.tex
\section{Problem Formulation}
\label{sec:problem}

\subsection{Preliminaries and Notation}
\label{sec:problem-prelim}
We consider a long-term memory backend over $T$ turns. At turn
$t\in\{1,\ldots,T\}$, the backend maps an incoming event $e_t$ and current
snapshot $M_t$ to $M_{t+1}$. This transition is not monolithic: it
contains latent state-evolution choices that are semantically admissible
but not uniquely determined.

We formalize each such choice as a memory-evolution decision
\begin{equation}
\begin{aligned}
\mathrm{decision}_{t} = \langle\,& \mathrm{carrier}_{t},\ \mathrm{candset}_{t},\\
& \mathrm{probdist}_{t},\ \mathrm{ctx}_{t} \rangle
\end{aligned}
\end{equation}
where $\carrier_t$ is the carrier type,
$\candset_{t}=\{c_{t}^{1},\ldots,c_{t}^{k_{t}}\}$ is the candidate set,
$\probdist_t\in\Delta(\candset_t)$ is the backend preference distribution,
and $\ctx_t$ is reconstructible context.

We assume the backend induces a latent preference distribution
$\probdist_t^\star$ over $\candset_t$, and that the baseline backend would
commit $c_t\sim\probdist_t^\star$. We elicit an explicit estimate
$\probdist_t$ from self-reported model weights and define distribution
preservation with respect to this estimate:
\begin{equation}
\hat c_t \sim \probdist_t \quad \text{for all } t\in\{1,\ldots,T\}.
\end{equation}

\paragraph{Decision--commit separation.}
\memmark{} observes and modifies only $(\candset_t, \probdist_t, \ctx_t)$;
candidate generation and the commit step
$\textsc{apply\_selected}(M_t,\hat c_t)\to M_{t+1}$ remain on the native
backend path. Thus, the watermark changes only the selected candidate, not
the backend write operation itself.

\subsection{Memory-Evolution Channel and Distribution-Preserving Coding}
\label{sec:problem-channel}
We model each decision stage as a time-varying discrete channel. To embed
a provenance payload, \memmark{} replaces native sampling with a
distribution-preserving encoder:
\begin{equation}
\hat c_t \leftarrow \mathsf{Enc}(\probdist_t, r_t)\in\candset_t,
\end{equation}
where $r_t$ is per-decision randomness reproducible by the embedder and
verifier.

\paragraph{Keyed pseudorandomness.}
The watermark secret $\wmkey$ and context derive a nonce
\begin{equation}
\nonceT \leftarrow \mathrm{PRF}(\wmkey,\ \ctx_t),
\end{equation}
which seeds the sampling stream consumed by $\mathsf{Enc}$. Since
$\ctx_t$ is reconstructible from surviving evidence, the verifier can
replay the keyed pick.

\subsection{Threat Model}
\label{sec:problem-threat}
The watermark is embedded at write time but verified later, possibly
after backend maintenance and partial or full loss of the external audit
log.

\paragraph{Structural attacks on the backend.}
Memory stores may be poisoned or compacted~\cite{chen2024agentpoison,
amemguard,memorygraft}. Such operations can remove, merge, or rewrite
records, but cannot forge a commitment that opens against the anchored
Merkle root without $K$. We therefore model the dominant effect as a
surviving decision set $\mathcal{I}\subseteq\{1,\ldots,T\}$.

\paragraph{Verification regimes.}
We consider three deployment regimes:
\begin{itemize}
  \item \textbf{R1 (full external log):} the verifier holds all
  per-decision commitments and the complete Merkle tree.
  \item \textbf{R2 (partial external log):} only a subset $\mathcal{I}$ of
  commitments survives, due to truncation, retention limits, or partial
  loss.
  \item \textbf{R3 (snapshot only):} the external log is unavailable;
  verification relies on the memory snapshot plus an in-record sidecar
  carrying reveal data and the anchored header.
\end{itemize}

\subsection{Objectives}
\label{sec:problem-objectives}
We formalize two objectives under the hard constraint
$\hat c_t\sim\probdist_t$.

\paragraph{Utility preservation.}
Watermarking should preserve downstream memory quality. For baseline and
watermarked snapshots $M_T$ and $\hat M_T$, we require
\begin{equation}
\begin{aligned}
\Big|\mathbb{E}[\mathcal{U}(\hat M_T)] - \mathbb{E}[\mathcal{U}(M_T)]\Big|
\le \varepsilon_{\mathcal{U}},
\end{aligned}
\end{equation}
where $\mathcal{U}(\cdot)$ is any downstream utility metric. This
motivates carriers whose candidates are semantically equivalent.

\paragraph{Robust attribution.}
The verifier should recover provenance whenever enough decisions survive.
In R1 and R2, given $\mathcal{I}$, we require
\begin{equation}
\begin{aligned}
\Pr\!\left[
\mathsf{Verify}\big(\{\hat c_t,\probdist_t,\ctx_t\}_{t\in\mathcal{I}};\ K\big)=1
\right] \\
\ge 1-\delta(|\mathcal{I}|),
\end{aligned}
\end{equation}
with failure probability $\delta(|\mathcal{I}|)$ decreasing as
$|\mathcal{I}|$ grows. R3 uses the in-record sidecar instead of external
commitments.

\section{MemMark}
\label{sec:method}

\subsection{Overview}
\label{sec:method-overview}
Figure~\ref{fig:pipeline} gives the roadmap. \memmark{} exposes
carrier-specific choices in a backend, turns them into a discrete
candidate distribution at the LLM-call boundary, selects $c^*$ with a
secret-keyed sampler, and stores the evidence needed for R1--R3
verification. We follow the same structure below: carriers and adapter
hooks (Section~\ref{sec:method-carrier}), distribution-preserving
sampling (Section~\ref{sec:method-sampler}), and the cryptographic audit
trace (Section~\ref{sec:method-audit}).

\subsection{Carrier Taxonomy and Backend Adapter}
\label{sec:method-carrier}
\input{tables/carrier_taxonomy}

Table~\ref{tab:carrier-taxonomy} lists our three carriers. The
update-target carrier changes which existing object is modified; the
link-target carrier changes which prior object is connected to new
evidence; and the semantic-realization carrier changes the surface form
encoding the same fact. Each captures backend decision freedom while
satisfying the semantic-equivalence requirement of
Section~\ref{sec:problem-objectives}.

Connecting a backend to \memmark{} requires three adapter hooks:
\textsc{enumerate\_candidates}$(M_t,e_t,\carrier)\to\candset$,
\textsc{score\_candidates}$(\candset,\ctx)\to\probdist$, and
\textsc{apply\_selected}$(M_t,c^*)\to M_{t+1}$. They expose the choice
space, score candidates, and commit the selected candidate. The same
abstraction also yields the carrier-level entropy and payload-allocation
statistics used in RQ2. Since the
adapter is the only backend-specific code path, one watermarking layer can
operate over structurally distinct systems such as \amem{} and
\graphiti{}.

\subsection{Distribution-Preserving Behavioral Watermark}
\label{sec:method-sampler}

\paragraph{Decision interception and elicitation.}
Each internal LLM call provides an intervention boundary: the backend
would often accept several nearby alternatives without changing the
semantic role of the write. \memmark{} extends the prompt to request $K$
plausible answers with self-reported preference weights. Parsing this
wrapper response yields candidates $\{d_t^1,\ldots,d_t^K\}$ and a
normalized distribution $\probdist_t$, reducing open-vocabulary
generation to a discrete choice that can be keyed, audited, and replayed.
The backend receives only the final selected response.

\paragraph{Keyed distribution-preserving sampling.}
\memmark{} feeds $(\candset_t,\probdist_t,\ctx_t)$ to a
distribution-preserving binning sampler keyed by the watermark secret $\wmkey$ and the
context-bound nonce $\nonceT$ from Section~\ref{sec:problem-channel}. The
sampler returns $\hat c_t$ and the number of embedded bits while preserving
the marginal:
\begin{equation}
\Pr[\hat c_t=c\mid \probdist_t,\wmkey]=\probdist_t(c),\quad \forall c\in\candset_t.
\end{equation}

\paragraph{Properties.}
The sampler provides the following guarantees; proofs are deferred to
Appendix~\ref{app:proofs}.

\begin{lemma}[Strict distribution preservation]
\label{lem:dist-preserve}
For any valid $\probdist_t$, the keyed pick $\hat c_t$ satisfies
$\Pr[\hat c_t=c^i\mid \wmkey,\ctx_t]_{\mathrm{marg}}=\probdist_t(c^i)$.
\end{lemma}

\begin{lemma}[Cascade composition]
\label{lem:cascade}
Keyed picks across multiple calls of a single event are independent
under distinct context-derived nonces, so expected embedding capacity
grows approximately additively in the number of valid decisions.
\end{lemma}

\begin{lemma}[Backend invariance]
\label{lem:backend-invariant}
Once $(\candset_t,\probdist_t,\ctx_t)$ is fixed, the keyed pick depends
only on the candidate distribution and secret key, not on the backend that
produced the candidates.
\end{lemma}

\subsection{Cryptographic Audit Trace}
\label{sec:method-audit}
To support verification after execution, \memmark{} records a
cryptographic trace for the regimes in Section~\ref{sec:problem-threat}.

\paragraph{Per-decision commitments.}
Each sampled decision produces a commitment
\[
\begin{aligned}
\commit_t = H\big(\,
& \ctx_t \,\|\, H(\candset_t) \,\|\, H(\probdist_t) \,\|\, \\
& \hat c_t \,\|\, \text{bits}_t \,\|\, \nonceT \,\big),
\end{aligned}
\]
with $H$ collision-resistant and inputs canonically serialized.

\paragraph{Per-session Merkle tree and signed anchor.}
Per-decision commitments form a Merkle tree whose root is sealed in
\[
\begin{aligned}
\header = (\,&\text{agent\_id},\ \text{user\_id},\ \text{session\_id},\ T,\\
            &\text{root}_T,\ \mathrm{sig}_{\wmkey}(\text{root}_T)),
\end{aligned}
\]
making the trace tamper-evident.

\paragraph{Per-leaf inclusion proofs.}
Each record carries its Merkle inclusion path at seal time, so a surviving
leaf can be checked directly against the anchored root without rebuilding
the log. Thus, structural attacks do not collapse verification into an
all-or-nothing rebuilt-root test.

%% file: tables/carrier_taxonomy.tex
\begin{table}[t]
\centering
\small
\setlength{\tabcolsep}{0.2mm}
\begin{tabular}{lll}
\toprule
Carrier $\carrier$ & \amem{} & \graphiti{}  \\
\midrule
\texttt{update\_target}
  & note id 
  & fact-edge id \\
\texttt{link\_target}
  & keyword cluster 
  & entity attach point \\
\texttt{semantic\_realize.}
  & note description 
  & edge label  \\
\bottomrule
\end{tabular}
\caption{\textbf{Carriers Taxonomy}. Each $\carrier$ is a
non-trivial-candidate-set evolve decision realized in
backend-specific form; the watermark sampler reads only
$(\candset, \probdist, \ctx)$ and is therefore backend-invariant.}
\label{tab:carrier-taxonomy}
\end{table}

%% file: sections/experiments.tex
%
%
\section{Experiments}
\label{sec:exp}
We evaluate utility preservation (RQ1), capacity (RQ2), snapshot-only
verification (RQ3), robustness to memory-lifecycle attacks (RQ4), and
memory integrity (RQ5).

\subsection{Setup}
\label{sec:setup}
\paragraph{Backends \& Models.}
We test two long-term memory backends: \amem{} \cite{amem}, a
note-centric system with dynamic linking and memory evolution, and
\graphiti{} \cite{graphiti}, a temporal graph memory with evolving
entities and relations. We use Deepseek-V4-pro
\cite{deepseekai2026deepseekv4}, Qwen3.6-flash \cite{qwen36_35b_a3b},
and GLM-5 \cite{glm5team2026glm5vibecodingagentic}. Candidate
enumeration uses $K{=}4$ and $T_{\text{enum}}{=}0.7$ by default
(Appendix~\ref{app:hyperparam}), candidate scoring uses
$T_{\text{score}}{=}0.0$, and JSON mode is enabled throughout.

\paragraph{Benchmark.}
We use \locomo{} \cite{maharana-etal-2024-evaluating}, which contains
ten multi-session dialogues with QA annotations spanning single-hop,
multi-hop, temporal, commonsense, and adversarial.

\paragraph{Baselines.}
We compare \memmark{} with four controls: \texttt{no-watermark},
\texttt{random-replace}, \texttt{signed-metadata-only}, and \kgmark{}
\cite{kgmark} on \graphiti{}, isolating native utility, unkeyed
randomization, signed metadata without embedding, and the closest
structured-memory watermark baseline.

\paragraph{Metrics.}
Appendix~\ref{app:metrics} defines all metrics. RQ1 reports \locomo{}
F1, BLEU-1, and F1 deltas; RQ2 reports per-carrier entropy, payload
share, and bits per decision; RQ3 reports R1/R2/R3 bit recovery and
wrong-key bit recovery; RQ4 reports post-attack recovery, the verifier
mode triggered by each attack, and wrong-key separation above the baseline
under nine attacks\footnote{Attack definitions are in
Appendix~\ref{app:attack}.}; and RQ5 reports carrier distribution,
evidence-grounded retrieval recall, and write failures.

\subsection{RQ1 --- Utility Preservation}
\label{sec:rq1}
\input{tables/main_results}

Table~\ref{tab:main-results} compares \memmark{} with unwatermarked and
ablated controls on \locomo{}. Across the six model--backend settings,
\memmark{} preserves downstream utility: the average Overall F1 changes
from $0.2816$ for \texttt{no-watermark} to $0.2804$ for \memmark{},
retaining $99.6\%$ of the unwatermarked score, while average BLEU-1
changes from $0.3069$ to $0.3077$, retaining $100.2\%$. The absolute
mean F1 drop is only $0.0012$, and BLEU-1 increases slightly by
$0.0008$.

The effect differs by backend but remains small. On \amem{}, \memmark{}
improves average Overall F1 from $0.3141$ to $0.3255$ and BLEU-1 from
$0.3420$ to $0.3529$. On \graphiti{}, average Overall F1 decreases from
$0.2490$ to $0.2353$, and BLEU-1 decreases from $0.2718$ to $0.2624$.
Thus, the watermark does not introduce a systematic utility collapse:
the attribution signal is embedded in ordinary memory-evolution decisions
while keeping task performance close to the native systems.

The comparison with \texttt{signed-metadata-only} is especially useful:
signing explicit provenance fields leaves utility largely unchanged, but
does not create a recoverable state-evolution signal. \memmark{} remains
in a comparable utility regime while moving the attribution evidence into
the backend's own choices.
\begin{figure}[!t]
\centering
\includegraphics[width=\linewidth]{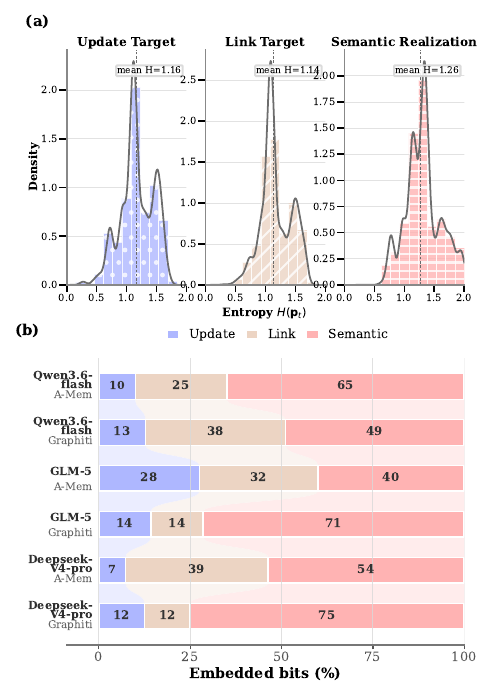}
\caption{\textbf{RQ2 --- Entropy and payload allocation across
memory-evolution carriers.}
(a) Per-carrier entropy over the six LLM--backend configurations; labels
show means. (b) Per-carrier share of embedded payload bits, normalized
within each configuration.}
\label{fig:rq2-combined}
\end{figure}
\subsection{RQ2 --- Capacity}
\label{sec:rq2}
\begin{figure}[!t]
\centering
\includegraphics[width=\linewidth]{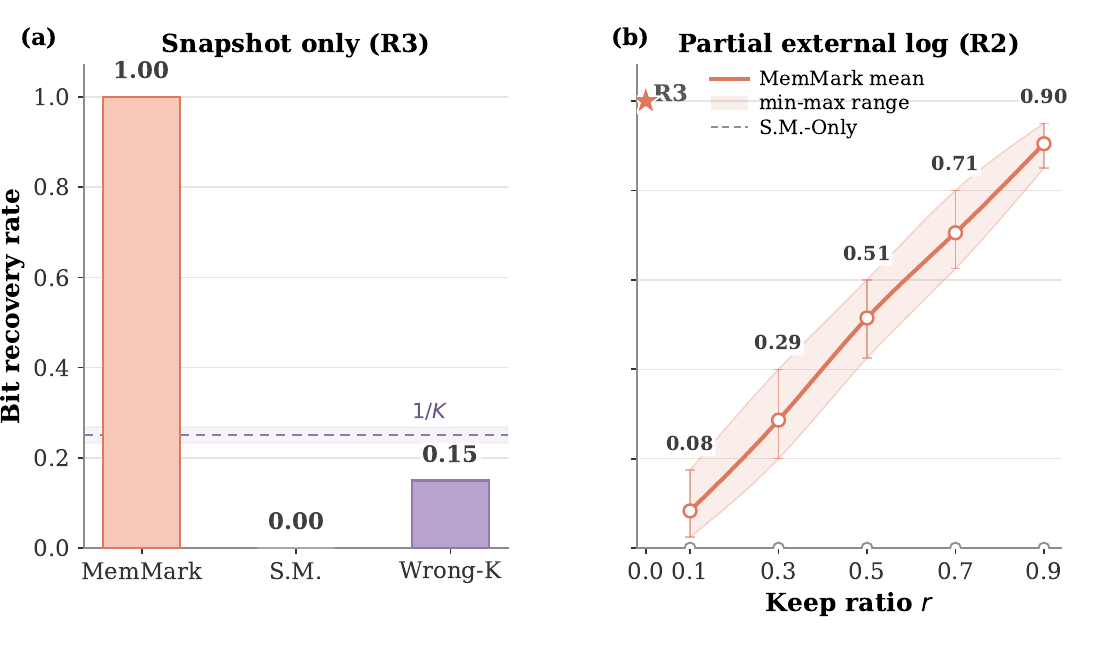}
\caption{\textbf{R3 snapshot verification and R2 partial-log degradation.}
Left: \memmark{} recovers all bits from the snapshot, while
\texttt{signed-metadata-only} fails and the wrong-key control remains at
chance level. Right: mean R2 recovery over six LLM--backend settings;
shading and error bars show the minimum--maximum range.}
\label{fig:r2-curve}
\end{figure}
Figure~\ref{fig:rq2-combined} shows non-trivial capacity across all three
carriers. The mean per-decision entropies are $1.16$, $1.14$, and
$1.26$ bits for update-target, link-target, and semantic-realization
decisions, respectively. Thus, each carrier exposes roughly one bit of
usable decision entropy per eligible write, and semantic realization is
the highest-capacity carrier average.

Capacity follows backend structure rather than a fixed carrier recipe.
For example, \amem{} distributes payload more evenly across carriers:
with Qwen3.6-flash, the payload split is $10\%$ update, $25\%$ link, and
$65\%$ semantic; with GLM-5, it becomes $28\%$, $32\%$, and $40\%$.
By contrast, \graphiti{} places more payload on semantic realization:
semantic decisions carry $49\%$--$75\%$ of embedded bits across the three
LLM backbones. The signal is therefore distributed over natural backend
choices rather than forced into one artifact.

This distribution is important for portability. \memmark{} does not
require two backends to expose the same internal objects; it only requires
each backend to present admissible choices at write time. The carrier mix
can change while the sampling and verification logic remains the same.

\subsection{RQ3 --- Snapshot-Only / Partial-Log Verification}
\label{sec:rq3}
RQ3 compares full-log verification (R1), partial logs with keep ratios
$r \in \{0.1,0.3,0.5,0.7,0.9\}$ (R2), and snapshot-only verification from
in-record reveal evidence and the signed session anchor (R3).

\input{tables/attack_signals}
\input{tables/integrity}

Figure~\ref{fig:r2-curve} shows that R3 matches full-log verification in
the benign setting. \memmark{} recovers the full 40-bit payload from the
final snapshot, giving a bit recovery rate of $1.00$. In contrast,
\texttt{signed-metadata-only} recovers $0.00$, and the wrong-key control
recovers only $0.15$, near the $1/K$ chance level.
Partial-log verification degrades smoothly with the amount of retained
evidence. As the keep ratio increases from $r=0.1$ to $0.9$, mean R2 bit
recovery rises from $0.08$ to $0.29$, $0.51$, $0.71$, and finally
$0.90$. This monotonic trend shows that verification is not an
all-or-nothing artifact of the complete log: even partial evidence
provides proportional attribution signal.

The gap to \texttt{signed-metadata-only} is the central RQ3 signal:
metadata can authenticate a cooperative writer, but it does not bind the
final memory state to latent evolution choices once the external log is
gone. In-record reveal evidence and the signed session anchor preserve
enough replay material for snapshot-only attribution.
\subsection{RQ4 --- Robustness}
\label{sec:rq4}
We stress-test nine lifecycle attacks at strengths $0.1$, $0.3$, and $0.5$,
spanning content edits, removals, and synthesis-style mutations. We report
post-attack recovery (Rec), the dominant verifier mode (Mode), and
wrong-key separation
($\Delta\mathrm{WK}=\mathrm{Rec}-\mathrm{WrongKey}$).

Table~\ref{tab:attack-signal-split} shows that non-removal attacks remain
recoverable at mild strengths but diverge at strength $0.5$. On \amem{},
content edits reduce average Rec from $0.896$ to $0.472$ as strength
grows, while synthesis-style attacks are less destructive on average,
ending at Rec $0.602$. Additive poisoning is the mildest non-removal case
at strength $0.5$ (Rec $0.725$, $\Delta\mathrm{WK}=+0.493$), whereas
subgraph reanchoring is strongest, nearly reaching the wrong-key baseline
(Rec $0.346$, $\Delta\mathrm{WK}=+0.115$). Removal attacks behave
differently: pruning and deduplication preserve surviving authenticated
records, keeping Rec at $1.00$ and producing \texttt{Miss}, while edits,
compaction, and poisoning trigger \texttt{ComF}.

\subsection{RQ5 --- Memory Integrity}
\label{sec:rq5}

Finally, RQ5 checks whether watermarking materially changes the memory
write path. Table~\ref{tab:integrity} reports three \locomo{} probes:
carrier allocation, evidence-grounded retrieval recall, and write-path
failures. Overall, \memmark{} remains within the normal write regime:
\amem{} retrieval recall retains $89.3\%$ of the \texttt{no-watermark}
baseline, with write failures changing from $4$ to $3$; \graphiti{} recall
retains $97.5\%$, with only a small increase in write failures. Carrier
use is also not concentrated in one decision type: with Qwen, semantic
carriers account for $55.0\%$ of carrier decisions on \amem{} and
$58.6\%$ on \graphiti{}, consistent with RQ2. These probes suggest that
\memmark{} preserves snapshot verifiability without large observable
write-path disruptions.

%% file: tables/main_results.tex
\definecolor{QwenPanel}{RGB}{226,240,220}
\definecolor{GlmPanel}{RGB}{226,236,246}
\definecolor{DeepseekPanel}{RGB}{248,229,232}
\newcommand{\rqbest}[1]{\textbf{#1}}
\newcommand{\rqsecond}[1]{\underline{#1}}
\newcommand{\modelpanel}[2]{\cellcolor{#1!30}\multirow{-9}{*}{\sffamily #2}}
\newcommand{\backendfour}[2]{\cellcolor{#1!34}\multirow{-4}{*}{#2}}
\newcommand{\backendfive}[2]{\cellcolor{#1!42}\multirow{-5}{*}{#2}}

\begin{table*}[!t]
  \centering
  \small
  \setlength{\tabcolsep}{3pt}
  \resizebox{\textwidth}{!}{%
  \begin{tabular}{lllcccccccccccc}
  \toprule
  & & & \multicolumn{2}{c}{Single Hop (1)} & \multicolumn{2}{c}{Temporal (2)} & \multicolumn{2}{c}{Multi Hop (3)} & \multicolumn{2}{c}{Open Domain (4)} & \multicolumn{2}{c}{Adversarial (5)} & \multicolumn{2}{c}{\textbf{Overall}} \\
  \cmidrule(lr){4-5}\cmidrule(lr){6-7}\cmidrule(lr){8-9}\cmidrule(lr){10-11}\cmidrule(lr){12-13}\cmidrule(lr){14-15}
  Model & Backend & Method & F1 & BLEU & F1 & BLEU & F1 & BLEU & F1 & BLEU & F1 & BLEU & F1 & BLEU \\
  \midrule
  \rowcolor{QwenPanel!18}
    & & \texttt{No-WM} & 0.1974 & 0.2373 & 0.3732 & 0.5360 & 0.1363 & 0.2333 & 0.4107 & 0.4265 & 0.1289 & 0.1234 & 0.2850 & 0.3322 \\
  \rowcolor{QwenPanel!18}
    & & \texttt{S.M.-Only} & 0.2127 & 0.2682 & \rqbest{0.4727} & \rqbest{0.5676} & \rqsecond{0.1884} & \rqsecond{0.3295} & \rqsecond{0.4115} & \rqsecond{0.4278} & \rqbest{0.2251} & \rqbest{0.2072} & \rqbest{0.3323} & \rqbest{0.3696} \\
  \rowcolor{QwenPanel!18}
    & & \texttt{Ran.} & \rqsecond{0.2305} & \rqbest{0.3364} & \rqsecond{0.4309} & \rqsecond{0.5541} & \rqbest{0.2595} & \rqbest{0.3761} & 0.3479 & 0.3810 & \rqsecond{0.1980} & \rqsecond{0.1859} & 0.3033 & \rqsecond{0.3596} \\
  \rowcolor{QwenPanel!18}
    & \backendfour{QwenPanel}{A-MEM} & \textbf{MemMark} & \rqbest{0.2559} & \rqsecond{0.3197} & 0.4143 & 0.5248 & 0.1840 & 0.2922 & \rqbest{0.4117} & \rqbest{0.4434} & 0.1243 & 0.1114 & \rqsecond{0.3044} & 0.3504 \\
  \rowcolor{QwenPanel!26}
    & & \texttt{No-WM} & \rqbest{0.1903} & \rqbest{0.2736} & 0.2867 & 0.3501 & 0.2905 & 0.3243 & 0.3543 & 0.3748 & \rqbest{0.1257} & \rqbest{0.1277} & \rqsecond{0.2572} & 0.2923 \\
  \rowcolor{QwenPanel!26}
    & & \texttt{S.M.-Only} & 0.1524 & 0.2131 & \rqsecond{0.3022} & 0.3604 & \rqsecond{0.3507} & \rqsecond{0.4615} & 0.3570 & \rqsecond{0.4023} & \rqsecond{0.1257} & \rqsecond{0.1277} & \rqbest{0.2589} & \rqbest{0.3031} \\
  \rowcolor{QwenPanel!26}
    & & \texttt{Ran.} & \rqsecond{0.1792} & \rqsecond{0.2587} & 0.3019 & \rqsecond{0.3658} & 0.3123 & 0.4231 & \rqbest{0.3614} & 0.3839 & 0.0488 & 0.0426 & 0.2440 & 0.2823 \\
  \rowcolor{QwenPanel!26}
    & & \texttt{KGMARK} & 0.1538 & 0.2365 & 0.2878 & 0.3536 & \rqbest{0.3845} & \rqbest{0.5556} & \rqsecond{0.3589} & \rqbest{0.4088} & 0.0887 & 0.0798 & 0.2505 & \rqsecond{0.3027} \\
  \rowcolor{QwenPanel!26}
  \modelpanel{QwenPanel}{Qwen3.6-flash}
    & \backendfive{QwenPanel}{Graphiti} & \textbf{MemMark} & 0.1389 & 0.2319 & \rqbest{0.4084} & \rqbest{0.4865} & 0.1865 & 0.3189 & 0.3364 & 0.3719 & 0.0701 & 0.0638 & 0.2453 & 0.2945 \\
  \midrule
  \rowcolor{GlmPanel!18}
    & & \texttt{No-WM} & 0.2340 & 0.2868 & 0.3562 & 0.3908 & \rqsecond{0.2319} & \rqsecond{0.2452} & \rqbest{0.4677} & \rqsecond{0.4592} & 0.0520 & 0.0438 & 0.2958 & 0.3067 \\
  \rowcolor{GlmPanel!18}
    & & \texttt{S.M.-Only} & \rqsecond{0.2342} & \rqsecond{0.3152} & \rqsecond{0.4144} & \rqbest{0.4689} & 0.1927 & 0.2088 & 0.4463 & \rqbest{0.4621} & 0.0408 & 0.0276 & 0.2939 & \rqsecond{0.3205} \\
  \rowcolor{GlmPanel!18}
    & & \texttt{Ran.} & 0.1875 & 0.2714 & \rqbest{0.4266} & 0.4428 & 0.1838 & 0.2001 & \rqsecond{0.4516} & 0.4571 & \rqbest{0.0710} & \rqbest{0.0643} & \rqsecond{0.2971} & 0.3150 \\
  \rowcolor{GlmPanel!18}
    & \backendfour{GlmPanel}{A-MEM} & \textbf{MemMark} & \rqbest{0.3067} & \rqbest{0.3491} & 0.4139 & \rqsecond{0.4438} & \rqbest{0.2774} & \rqbest{0.2803} & 0.4489 & 0.4491 & \rqsecond{0.0670} & \rqsecond{0.0638} & \rqbest{0.3181} & \rqbest{0.3300} \\
  \rowcolor{GlmPanel!26}
    & & \texttt{No-WM} & \rqbest{0.1686} & \rqbest{0.2731} & \rqbest{0.2844} & \rqsecond{0.3033} & 0.1474 & 0.1611 & 0.3487 & 0.3492 & 0.0914 & 0.0851 & \rqsecond{0.2338} & \rqsecond{0.2538} \\
  \rowcolor{GlmPanel!26}
    & & \texttt{S.M.-Only} & 0.1352 & 0.2203 & 0.2575 & 0.2713 & \rqsecond{0.1585} & 0.1519 & 0.3397 & \rqsecond{0.3633} & 0.0779 & 0.0673 & 0.2179 & 0.2395 \\
  \rowcolor{GlmPanel!26}
    & & \texttt{Ran.} & 0.1180 & 0.2154 & \rqsecond{0.2696} & \rqbest{0.3254} & 0.1567 & \rqsecond{0.1628} & 0.3390 & 0.3501 & 0.0488 & 0.0426 & 0.2101 & 0.2390 \\
  \rowcolor{GlmPanel!26}
    & & \texttt{KGMARK} & \rqsecond{0.1473} & \rqsecond{0.2324} & 0.2436 & 0.2550 & \rqbest{0.2237} & \rqbest{0.2643} & \rqbest{0.3674} & \rqbest{0.3773} & \rqbest{0.1126} & \rqbest{0.1064} & \rqbest{0.2394} & \rqbest{0.2599} \\
  \rowcolor{GlmPanel!26}
  \modelpanel{GlmPanel}{GLM-5}
    & \backendfive{GlmPanel}{Graphiti} & \textbf{MemMark} & 0.1060 & 0.1901 & 0.2518 & 0.2725 & 0.1199 & 0.1203 & \rqsecond{0.3532} & 0.3580 & \rqsecond{0.0968} & \rqsecond{0.0946} & 0.2188 & 0.2374 \\
  \midrule
  \rowcolor{DeepseekPanel!18}
    & & \texttt{No-WM} & \rqsecond{0.3275} & \rqbest{0.3888} & \rqsecond{0.4571} & \rqsecond{0.5476} & \rqsecond{0.2639} & 0.2912 & \rqbest{0.4899} & \rqbest{0.4948} & 0.1458 & 0.1254 & \rqsecond{0.3616} & \rqsecond{0.3870} \\
  \rowcolor{DeepseekPanel!18}
    & & \texttt{S.M.-Only} & 0.2768 & 0.3425 & \rqbest{0.4769} & \rqbest{0.5743} & \rqbest{0.3052} & \rqbest{0.3911} & \rqsecond{0.4873} & \rqsecond{0.4909} & 0.1632 & 0.1477 & \rqbest{0.3630} & \rqbest{0.3950} \\
  \rowcolor{DeepseekPanel!18}
    & & \texttt{Ran.} & 0.2105 & 0.2315 & 0.4359 & 0.5108 & 0.2335 & 0.3049 & 0.4605 & 0.4620 & \rqbest{0.2083} & \rqbest{0.1883} & 0.3413 & 0.3591 \\
  \rowcolor{DeepseekPanel!18}
    & \backendfour{DeepseekPanel}{A-MEM} & \textbf{MemMark} & \rqbest{0.3333} & \rqsecond{0.3651} & 0.4279 & 0.4962 & 0.2597 & \rqsecond{0.3251} & 0.4679 & 0.4818 & \rqsecond{0.1667} & \rqsecond{0.1552} & 0.3541 & 0.3784 \\
  \rowcolor{DeepseekPanel!26}
    & & \texttt{No-WM} & 0.1276 & 0.1448 & \rqbest{0.3659} & \rqbest{0.4108} & \rqsecond{0.2616} & \rqsecond{0.2865} & 0.3436 & \rqbest{0.3588} & 0.1249 & 0.1048 & \rqsecond{0.2560} & \rqbest{0.2694} \\
  \rowcolor{DeepseekPanel!26}
    & & \texttt{S.M.-Only} & 0.1286 & 0.1769 & 0.2836 & 0.2998 & 0.2060 & 0.2105 & 0.3109 & 0.3162 & \rqbest{0.1626} & \rqsecond{0.1323} & 0.2346 & 0.2404 \\
  \rowcolor{DeepseekPanel!26}
    & & \texttt{Ran.} & \rqbest{0.1653} & \rqbest{0.1989} & \rqsecond{0.3575} & \rqsecond{0.4050} & 0.1775 & 0.1835 & \rqbest{0.3585} & \rqsecond{0.3555} & 0.1266 & 0.1039 & \rqbest{0.2606} & \rqsecond{0.2689} \\
  \rowcolor{DeepseekPanel!26}
    & & \texttt{KGMARK} & \rqsecond{0.1304} & \rqsecond{0.1807} & 0.3213 & 0.3482 & 0.1815 & 0.2707 & \rqsecond{0.3470} & 0.3493 & \rqsecond{0.1432} & \rqbest{0.1361} & 0.2484 & 0.2665 \\
  \rowcolor{DeepseekPanel!26}
  \modelpanel{DeepseekPanel}{DeepSeek-V4-Pro}
    & \backendfive{DeepseekPanel}{Graphiti} & \textbf{MemMark} & 0.1104 & 0.1366 & 0.3353 & 0.3579 & \rqbest{0.2748} & \rqbest{0.3334} & 0.3229 & 0.3330 & 0.1277 & 0.1176 & 0.2418 & 0.2552 \\
  \bottomrule
  \end{tabular}%
  }
  \caption{\textbf{RQ1 --- Utility preservation on \locomo{}}. Results are grouped by model and backend. \texttt{No-WM}, \texttt{S.M.-Only}, \texttt{Ran.}, and \texttt{KGMARK} denote unwatermarked execution, signed-metadata-only, random replacement, and \kgmark{} (Graphiti only); \textbf{MemMark} is the full method. Bold and underlined entries mark the best and second-best scores within each backend block. Averaged over the six model--backend settings, \memmark{} changes Overall
F1 from $0.2816$ to $0.2804$, retaining $99.6\%$ of the unwatermarked
baseline, and changes BLEU-1 from $0.3069$ to $0.3077$, improved by
$0.2\%$.}
  \label{tab:main-results}
  \end{table*}

%% file: tables/attack_signals.tex
\begin{table*}[!t]
\centering
\scriptsize
\setlength{\tabcolsep}{2.2pt}
\renewcommand{\arraystretch}{1.22}

\definecolor{mmContent}{RGB}{224,242,241}
\definecolor{mmRemoval}{RGB}{252,241,241}
\definecolor{mmSynth}{RGB}{250,219,221}
\definecolor{recA}{RGB}{106,184,223}
\definecolor{recB}{RGB}{147,205,229}
\definecolor{recC}{RGB}{197,224,239}
\definecolor{recD}{RGB}{246,236,241}
\definecolor{recE}{RGB}{246,198,215}
\definecolor{recF}{RGB}{239,124,162}

\newcommand{\rA}[1]{\cellcolor{recA}\textbf{#1}}
\newcommand{\rB}[1]{\cellcolor{recB}\textbf{#1}}
\newcommand{\rC}[1]{\cellcolor{recC}\textbf{#1}}
\newcommand{\rD}[1]{\cellcolor{recD}\textbf{#1}}
\newcommand{\rE}[1]{\cellcolor{recE}\textbf{#1}}
\newcommand{\rF}[1]{\cellcolor{recF}\textbf{#1}}
\newcolumntype{H}{>{\centering\arraybackslash}m{2.45em}}

\resizebox{\textwidth}{!}{%
\begin{tabular}{@{}>{\centering\arraybackslash}m{4.8em}>{\centering\arraybackslash}m{5.6em}>{\centering\arraybackslash}m{4.1em}
HHH@{\hspace{2pt}}HHH@{\hspace{7pt}}
HHH@{\hspace{2pt}}HHH@{\hspace{7pt}}
HHH@{\hspace{2pt}}HHH@{}}
\toprule
& & & \multicolumn{6}{c}{\textbf{Deepseek-V4-pro}} & \multicolumn{6}{c}{\textbf{Qwen3.6-flash}} & \multicolumn{6}{c}{\textbf{GLM-5}} \\
\cmidrule(lr){4-9}\cmidrule(lr){10-15}\cmidrule(lr){16-21}
\textbf{Family} & \textbf{Attack} & \textbf{Mode}
& \multicolumn{3}{c}{\textbf{Rec}} & \multicolumn{3}{c}{\textbf{$\Delta\mathrm{WK}$}}
& \multicolumn{3}{c}{\textbf{Rec}} & \multicolumn{3}{c}{\textbf{$\Delta\mathrm{WK}$}}
& \multicolumn{3}{c}{\textbf{Rec}} & \multicolumn{3}{c}{\textbf{$\Delta\mathrm{WK}$}} \\
\midrule

\cellcolor{mmContent}\textbf{Content} & \cellcolor{mmContent}\textbf{Content-1} & \cellcolor{mmContent}\texttt{ComF} & \rA{0.93} & \rC{0.78} & \rD{0.54} & \rA{+0.73} & \rB{+0.59} & \rD{+0.34} & \rA{0.95} & \rD{0.60} & \rE{0.45} & \rB{+0.60} & \rD{+0.25} & \rE{+0.10} & \rA{0.90} & \rC{0.75} & \rD{0.62} & \rA{+0.75} & \rB{+0.60} & \rC{+0.47} \\
\cellcolor{mmContent} & \cellcolor{mmContent}\textbf{Content-2} & \cellcolor{mmContent}\texttt{ComF} & \rB{0.83} & \rC{0.68} & \rD{0.54} & \rB{+0.63} & \rC{+0.49} & \rD{+0.34} & \rB{0.80} & \rD{0.57} & \rE{0.45} & \rC{+0.45} & \rE{+0.22} & \rE{+0.10} & \rB{0.82} & \rC{0.68} & \rE{0.35} & \rB{+0.67} & \rC{+0.53} & \rE{+0.20} \\
\cellcolor{mmContent} & \cellcolor{mmContent}\textbf{Content-3} & \cellcolor{mmContent}\texttt{ComF} & \rA{0.90} & \rB{0.83} & \rD{0.56} & \rA{+0.71} & \rB{+0.63} & \rD{+0.37} & \rA{0.90} & \rC{0.72} & \rD{0.50} & \rB{+0.55} & \rD{+0.38} & \rE{+0.15} & \rA{0.97} & \rC{0.78} & \rD{0.53} & \rA{+0.82} & \rB{+0.62} & \rD{+0.38} \\
\cellcolor{mmContent} & \cellcolor{mmContent}\textbf{Content-4} & \cellcolor{mmContent}\texttt{ComF} & \rA{0.98} & \rB{0.83} & \rD{0.63} & \rA{+0.78} & \rB{+0.63} & \rC{+0.44} & \rA{0.90} & \rB{0.82} & \rD{0.62} & \rB{+0.55} & \rC{+0.47} & \rD{+0.28} & \rB{0.85} & \rD{0.50} & \rF{0.25} & \rA{+0.70} & \rD{+0.35} & \rE{+0.10} \\
\cellcolor{mmContent} & \cellcolor{mmContent}\textbf{Content-5} & \cellcolor{mmContent}\texttt{ComF} & \rA{0.90} & \rC{0.71} & \rE{0.44} & \rA{+0.71} & \rC{+0.51} & \rE{+0.24} & \rA{0.90} & \rD{0.53} & \rE{0.35} & \rB{+0.55} & \rE{+0.18} & \rF{+0.00} & \rA{0.90} & \rD{0.60} & \rF{0.25} & \rA{+0.75} & \rC{+0.45} & \rE{+0.10} \\

\midrule
\cellcolor{mmRemoval}\textbf{Removal} & \cellcolor{mmRemoval}\textbf{Removal-1} & \cellcolor{mmRemoval}\texttt{Miss} & \rA{1.00} & \rA{1.00} & \rA{1.00} & \rA{+0.80} & \rA{+0.80} & \rA{+0.80} & \rA{1.00} & \rA{1.00} & \rA{1.00} & \rB{+0.65} & \rB{+0.65} & \rB{+0.65} & \rA{1.00} & \rA{1.00} & \rA{1.00} & \rA{+0.85} & \rA{+0.85} & \rA{+0.85} \\
\cellcolor{mmRemoval} & \cellcolor{mmRemoval}\textbf{Removal-2} & \cellcolor{mmRemoval}\texttt{Miss} & \rA{1.00} & \rA{1.00} & \rA{1.00} & \rA{+0.80} & \rA{+0.80} & \rA{+0.80} & \rA{1.00} & \rA{1.00} & \rA{1.00} & \rB{+0.65} & \rB{+0.65} & \rB{+0.65} & \rA{1.00} & \rA{1.00} & \rA{1.00} & \rA{+0.85} & \rA{+0.85} & \rA{+0.85} \\

\midrule
\cellcolor{mmSynth}\textbf{Synthesis} & \cellcolor{mmSynth}\textbf{Synth-1} & \cellcolor{mmSynth}\texttt{ComF} & \rB{0.88} & \rD{0.61} & \rD{0.51} & \rB{+0.68} & \rC{+0.41} & \rD{+0.32} & \rA{0.90} & \rC{0.68} & \rE{0.40} & \rB{+0.55} & \rD{+0.33} & \rF{+0.05} & \rA{0.90} & \rC{0.78} & \rD{0.53} & \rA{+0.75} & \rB{+0.62} & \rD{+0.38} \\
\cellcolor{mmSynth} & \cellcolor{mmSynth}\textbf{Synth-2} & \cellcolor{mmSynth}\texttt{ComF} & \rA{0.93} & \rB{0.85} & \rC{0.72} & \rA{+0.74} & \rB{+0.66} & \rC{+0.52} & \rA{0.93} & \rB{0.80} & \rC{0.71} & \rB{+0.58} & \rC{+0.45} & \rD{+0.36} & \rA{0.91} & \rB{0.82} & \rC{0.74} & \rA{+0.76} & \rB{+0.67} & \rB{+0.59} \\

\midrule
& & & \multicolumn{3}{c}{0.1\quad 0.3\quad 0.5} & \multicolumn{3}{c}{0.1\quad 0.3\quad 0.5}
& \multicolumn{3}{c}{0.1\quad 0.3\quad 0.5} & \multicolumn{3}{c}{0.1\quad 0.3\quad 0.5}
& \multicolumn{3}{c}{0.1\quad 0.3\quad 0.5} & \multicolumn{3}{c}{0.1\quad 0.3\quad 0.5} \\
\bottomrule
\end{tabular}}
\caption{\textbf{RQ4 -- Attack-specific recovery and wrong-key separation
on \amem{}.} Each row reports the attack family, a compact row alias
that maps to Table~\ref{tab:attack-defs}
(\textbf{Content-1--5} $\rightarrow$ attacks 1--5,
\textbf{Removal-1--2} $\rightarrow$ attacks 6--7, and
\textbf{Synth-1--2} $\rightarrow$ attacks 8--9), and the dominant
verifier mode
(\texttt{ComF} = \texttt{commitment\_fail};
\texttt{Miss} = \texttt{missing\_leaves}). For each model, \textbf{Rec}
is post-attack bit recovery and $\Delta\mathrm{WK}=\mathrm{Rec}-
\mathrm{WrongKey}$ compares recovery with that model's wrong-key R3
baseline at attack strengths $s\in\{0.1,0.3,0.5\}$. Positive
$\Delta\mathrm{WK}$ means the attacked snapshot retains more attribution
signal than an incorrect key.}
\label{tab:attack-signal-split}
\end{table*}

%% file: tables/integrity.tex
\begin{table}[!tbp]
\centering
\scriptsize
\setlength{\tabcolsep}{3.0pt}
\resizebox{\columnwidth}{!}{%
\begin{tabular}{llccc}
\toprule
Backend & Method & Carrier Dist. & Ev. Rec. & Write Fail \\
\midrule
\multirow{4}{*}{\amem{}}
& \texttt{no-watermark}         & 284:90:477 & 0.597 & 4 \\
& \texttt{signed-metadata-only} & 251:135:462 & 0.560 & 3 \\
& \texttt{random-replace}       & 168:205:471 & 0.504 & 2 \\
& \textbf{\memmark}             & 237:143:465 & 0.533 & 3 \\
\midrule
\multirow{4}{*}{\graphiti{}}
& \texttt{no-watermark}         & 288:481:834 & 0.236 & 0 \\
& \texttt{signed-metadata-only} & 274:422:917 & 0.198 & 4 \\
& \texttt{random-replace}       & 264:384:920 & 0.176 & 1 \\
& \textbf{\memmark}             & 278:386:938 & 0.230 & 5 \\
\bottomrule
\end{tabular}%
}
\caption{\textbf{RQ5 -- Memory-integrity probes on \locomo{}.}
All rows use the Qwen backbone. \textbf{Carrier Dist.} gives
update:link:semantic counts; \textbf{Ev. Rec.} is evidence-grounded
retrieval recall; \textbf{Write Fail} counts write-path failures.}
\label{tab:integrity}
\end{table}

%% file: sections/conclusion.tex
\section{Conclusion}
\label{sec:conclusion}

\memmark{} watermarks the state-evolution layer of long-term agent memory,
binding provenance to backend write, update, linking, and retention choices.
The results show that durable memory attribution can survive beyond visible
actions, final text, and trusted metadata when it is tied to
utility-preserving state choices and authenticated reveal evidence.

%% file: sections/limitation.tex
\section{Limitations}
\label{sec:limitations}
\memmark{} is evaluated on two memory backends and one auditable benchmark,
leaving broader deployment settings to future work. In particular, longer
memory lifecycles, backend-specific compaction, migration, and periodic
summarization may introduce new watermark carriers as well as new sources
of drift. Future work should extend the adapter contract to cover these
maintenance operations. The current attack study covers representative edits, deletions, and poisoning, but does not exhaust adaptive attempts to remove evidence while preserving utility. A natural next step is to study such attacks and build verification procedures that distinguish benign lifecycle changes from targeted tampering, making snapshot-only attribution more robust in production systems.

%% file: sections/appendix.tex

\section{Experiment Details}

\subsection{Data Statistics}
We evaluate on the public LoCoMo benchmark, using the same fixed set of ten long-term multi-session conversations for all main model--backend configurations. This fixed-conversation protocol ensures that comparisons across methods, memory backends, and LLM backbones are made on the same evaluation instances rather than on backend- or model-specific subsets.

LoCoMo contains 10 conversations. In the release, each conversation contains an average of 27.2 sessions, 21.6 turns per session, and 16,618.1 tokens per conversation. The average dialogue turn contains 29.8 tokens. LoCoMo also provides derived memory artifacts: observations average 19.2 tokens each, and session summaries average 132.4 tokens. The benchmark is multimodal as well, with an average of 91.2 images per conversation, although our experiments use LoCoMo for memory-oriented QA evaluation rather than multimodal generation.

For the QA benchmark, LoCoMo provides 1,986 total questions across five reasoning categories. These include 841 single-hop questions (42.3\%), 282 multi-hop questions (14.2\%), 321 temporal-reasoning questions (16.1\%), 96 open-domain knowledge questions (4.8\%), and 446 adversarial questions (22.4\%). Single-hop questions require evidence from one session, multi-hop questions require synthesizing information across multiple sessions, temporal questions require reasoning over time cues, open-domain questions require combining dialogue information with general knowledge, and adversarial questions are designed to be unanswerable or misleading.

\subsection{Artifact Attribution and Licenses}

We use these artifacts under their stated access conditions and licenses. The A-MEM reproduction repository is released under the MIT License, and GRAPHITI is released under the Apache License 2.0. For LoCoMo, we use the dataset and code for research evaluation according to the terms provided with the released repository. For model backbones, we access the models through their public model releases or APIs and follow the corresponding model licenses, platform terms, and usage policies. 

\subsection{Consistency with Intended Use}
Our use of the artifacts is consistent with their intended research purposes. LoCoMo is used only as an evaluation benchmark for long-term conversational memory. We do not use LoCoMo-derived data for deployment, user profiling, or non-research applications. A-MEM and GRAPHITI are used as memory-system substrates, matching their intended role as agent-memory frameworks. The LLM backbones are used as controlled model components for candidate generation, memory writing, and memory-based QA; they are not fine-tuned on LoCoMo, and we do not redistribute model weights beyond the access conditions of the original providers.

\subsection{Potential Risks}
MEMMARK introduces a provenance mechanism for long-term agent memory, but it also creates risks that must be considered in deployment. First, watermark verification depends on secret keys, canonicalized reveal evidence, and cryptographic audit material. If keys are leaked, sidecars are mishandled, or canonical serialization is inconsistent across systems, verification may become unreliable. Second, successful watermark verification should not be interpreted as a guarantee that the memory content is true or safe. It only shows that the observed memory evolution is consistent with the keyed sampler and audit trace.

\section{Full Proofs}
\label{app:proofs}

This appendix collects the technical material that supports the main
paper's central claims but would be too detailed for the main
narrative. We begin with proof sketches for the sampler properties,
then define the metrics and baselines before reporting the experimental
appendix results referenced in the main text.

\subsection{Proof of Lemma~\ref{lem:dist-preserve}}
Let $\candset_t=\{c^1,\ldots,c^k\}$ and write
$p_i=\probdist_t(c^i)$. The sampler uses the same integer-binning
argument as \agentmark{}~\cite{agentmark}. Choose an audit precision
$N$ and represent the serialized distribution by non-negative integer
masses $n_i$ with $\sum_i n_i=N$ and $n_i/N=p_i$. In practice
$\probdist_t$ is the canonical, finite-precision distribution stored in
the reveal record, so the equality is exact with respect to the
audited distribution.

Partition the cyclic group $\mathbb{Z}_N$ into consecutive intervals
$I_i$ with $|I_i|=n_i$, one interval per candidate. For any payload
position $x\in\mathbb{Z}_N$, the keyed sampler draws a shift
$s=\mathrm{PRF}(\wmkey,\ctx_t)\bmod N$ and selects the unique $c^i$ such
that $(x+s)\bmod N\in I_i$. In the ideal experiment $s$ is uniform on
$\mathbb{Z}_N$; under a secure PRF it is computationally
indistinguishable from uniform to an observer without $K$. Therefore
$(x+s)\bmod N$ is uniform regardless of the payload point $x$, and
\begin{align*}
\Pr[\hat c_t=c^i]
  &= \Pr[(x+s)\bmod N\in I_i] \\
  &= \frac{|I_i|}{N}
   = p_i .
\end{align*}
Thus the keyed pick has exactly the same marginal distribution as
self-reported sampling from $\probdist_t$. The notation
$\Pr[\cdot\mid \wmkey,\ctx_t]_{\mathrm{marg}}$ in the lemma should be read as
the standard watermarking marginal over the PRF-key experiment, or
equivalently over the pseudorandom shift that the hidden key induces.
After a concrete key and context are fixed, the selector is of course
deterministic and replayable by the verifier.

\subsection{Proof of Lemma~\ref{lem:cascade}}
Consider one memory event that triggers $m$ internal LLM calls. For
call $j$, the adapter exposes
$(\candset_{t,j},\probdist_{t,j},\ctx_{t,j})$, where
$\ctx_{t,j}$ contains the round index, dialogue identifiers, prompt
hash, and the previous commitment. These fields domain-separate the
calls: except with negligible collision probability,
$\ctx_{t,j}\neq\ctx_{t,\ell}$ for $j\neq \ell$.

The backend may choose later contexts adaptively after observing earlier
selected candidates. PRF security covers such adaptively chosen distinct
inputs, so the sequence
$\mathrm{PRF}(\wmkey,\ctx_{t,1}),\ldots,\mathrm{PRF}(\wmkey,\ctx_{t,m})$ is
computationally indistinguishable from independent uniform draws. By
Lemma~\ref{lem:dist-preserve}, each individual keyed pick preserves its
own marginal distribution:
\[
\Pr[\hat c_{t,j}=c\mid \candset_{t,j},\probdist_{t,j},\ctx_{t,j}]
=\probdist_{t,j}(c).
\]
Independence of the PRF outputs then gives the product-form joint
distribution conditioned on the exposed decision tuples. If $B_{t,j}$
is the number of payload bits embedded at decision $j$, total capacity
over the cascade is
$B_t=\sum_{j=1}^{m}B_{t,j}$ and hence
$\mathbb{E}[B_t]=\sum_{j=1}^{m}\mathbb{E}[B_{t,j}]$ by linearity of
expectation. Since every stage is marginally distribution preserving,
there is no accumulating sampling bias across the cascade; additional
calls add evidence and capacity, not systematic drift.

\subsection{Proof of Lemma~\ref{lem:backend-invariant}}
Fix any backend $\mathcal{B}$ that implements the adapter interface and
emits a valid decision tuple
$(\candset_t,\probdist_t,\ctx_t)$. The sampler receives only this tuple
and the watermark secret. It does not inspect whether the candidates are
\amem{} notes, \graphiti{} entities, graph edges, or surface
realizations. Applying Lemma~\ref{lem:dist-preserve} to the emitted tuple
therefore yields
\[
\Pr_{\mathcal{B}}[\hat c_t=c^i\mid \candset_t,\probdist_t,\ctx_t]
=\probdist_t(c^i),
\]
with no term depending on $\mathcal{B}$ beyond the tuple itself. If two
backends expose the same $(\candset_t,\probdist_t,\ctx_t)$, the keyed
sampler induces the same marginal law and the same replay rule. This
does not assert that all backends produce identical candidate
distributions; rather, it states that once a backend has reduced its
native write choice to the common adapter representation, the watermark
layer supplies the same distribution-preservation guarantee. This is
the invariant used by the cross-backend comparisons in the main text.

\section{Metrics}
\label{app:metrics}

Table~\ref{tab:metric-defs} consolidates the definitions of all metrics
used in this paper, grouped by the research question they address: utility
(RQ1), capacity (RQ2), snapshot-only verification (RQ3), tamper detection
(RQ4), and memory integrity (RQ5). For each metric we list its reporting
unit and, where relevant, the carrier-level decomposition over update
target, link target, and semantic realization. We refer the reader to
this table throughout Sections~\ref{sec:rq1}--\ref{sec:rq5} for precise
definitions.

\input{tables/metric_definitions}

\input{sections/baseline_details}

\section{Hyperparameter Sensitivity}
\label{app:hyperparam}
We perform a default-centered one-factor sensitivity analysis on
Qwen3.6-flash with the \amem{} backend. Starting from the default
$K{=}4,T_{\text{enum}}{=}0.7$, we vary
$K\in\{2,4,8\}$ while holding $T_{\text{enum}}{=}0.7$, and vary
$T_{\text{enum}}\in\{0.5,0.7,1.0\}$ while holding $K{=}4$. This isolates
the two expected trade-offs: candidate count controls the capacity and
wrong-key chance level, while enumeration temperature controls the
diversity--utility balance.

\input{tables/hyperparam_sweep}

Table~\ref{tab:hyperparam} shows that the default setting is a stable
middle point rather than a tuned extreme. Reducing $K$ to $2$ lowers the
realized capacity and pushes wrong-key recovery toward the $1/2$
chance-level reference. Increasing $K$ to $8$ improves bits per decision
and separates wrong-key recovery more strongly, but the utility delta
becomes more negative, consistent with lower-quality candidates entering
the admissible set. Temperature has a similar but softer effect:
$T_{\text{enum}}{=}0.5$ is conservative and slightly improves utility,
whereas $T_{\text{enum}}{=}1.0$ increases capacity at the cost of a
larger F1 drop. R3 recovery remains complete across the sweep, indicating
that these settings affect capacity and perturbation more than benign
snapshot verifiability.

\section{Overall and Per-Conversation Experimental Results}
\label{app:overall-per-conv}

\begin{figure}[H]
  \centering
  \includegraphics[width=0.95\columnwidth]{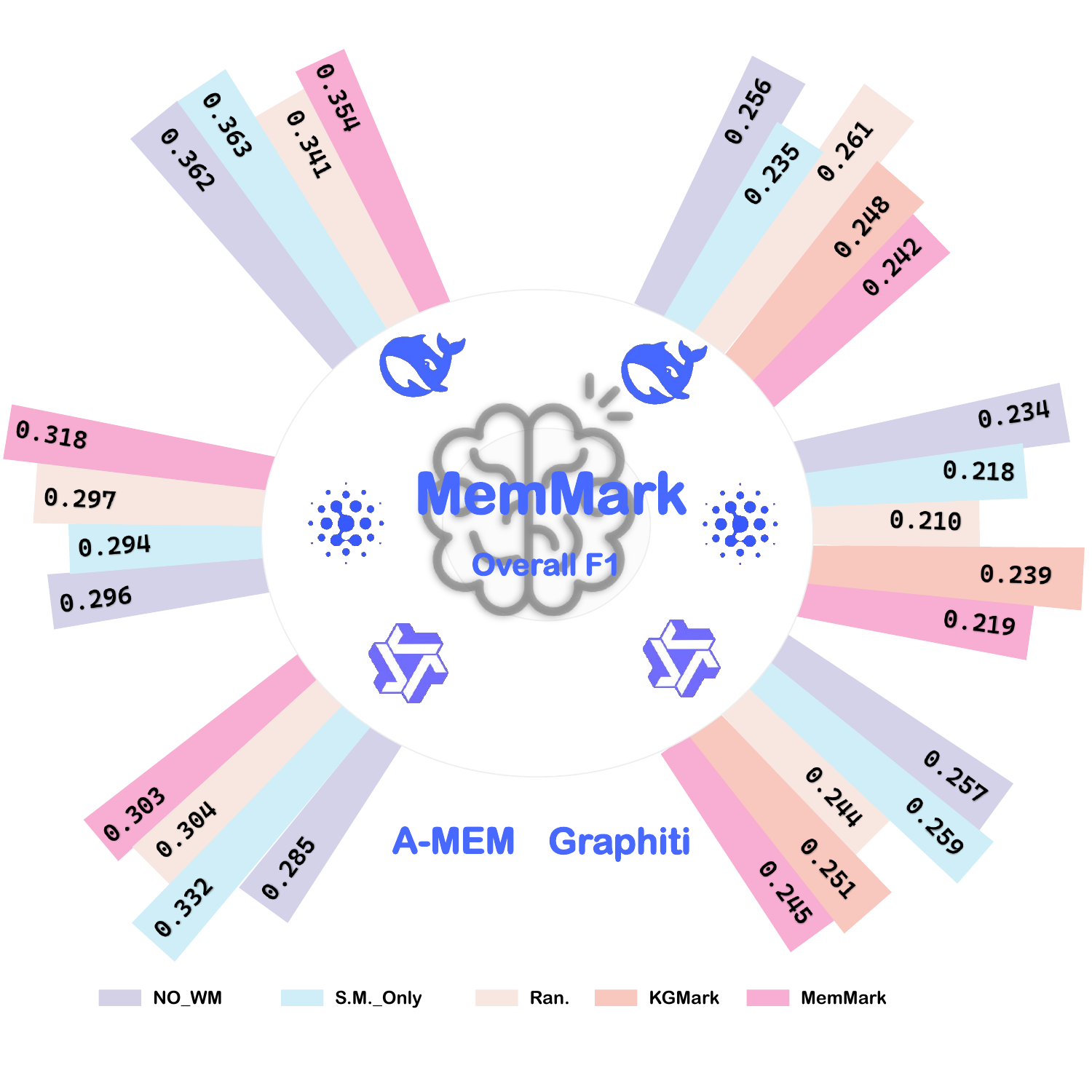}
  \caption{\textbf{Overall F1 comparison}. Overall F1 from
  Table~\ref{tab:main-results} across three LLMs and two memory
  backends, comparing \memmark{} with baselines.}
  \label{fig:overall-f1-app}
\end{figure}

Figure~\ref{fig:overall-f1-app} provides an appendix visualization of
the main experimental table, focusing on overall F1 across the three LLMs
and the two backends. We then provide the per-conversation breakdown for
Qwen3.6-flash on all 10 \locomo{} conversations. This table is included
only in the appendix because its role is diagnostic rather than
conceptual: it shows that the aggregate trends in the main text are not
driven by a single conversation outlier.

The overall plot makes two points visible at once. First, \memmark{}
tracks the main utility curve rather than collapsing it: across the three
LLMs, the watermarked runs remain in the same regime as their
unwatermarked counterparts. Second, the backend effect is larger than the
watermark effect. \amem{} is generally more stable under the watermarking
wrapper, while \graphiti{} shows a larger spread and a slightly stronger
drop in the hardest settings. That is exactly the pattern expected from a
memory system whose candidate space is backend-dependent but still
compatible with the same sampler.

\input{tables/per_conversation}

Table~\ref{tab:per-conversation} shows the same story at the
conversation level. The \amem{} deltas stay small, mostly within a few
points, and are centered close to zero, with a mean gap of $+0.006$.
\graphiti{} is more variable and more sensitive to conversation-specific
structure, but the mean gap remains modest at $-0.038$ over the
all-conversation Qwen3.6-flash diagnostic. In other words, the watermark does not
introduce a single brittle failure mode; it behaves like a mild shift on
top of ordinary conversational variance. The verification columns remain
stable across conversations: R1 and R3 recover the payload completely,
R2 at $r{=}0.5$ averages $0.556$, and the wrong-key control averages
$0.205$, close to the $1/K{=}0.25$ chance reference. Thus, the capacity
signal survives the same diversity that perturbs utility while remaining
key-specific.

\section{Memory-lifecycle Attacks and Backend Diagnostics}
\label{app:attack}

We evaluate robustness (RQ4) under nine attacks that span the realistic
lifecycle of an agent's memory store. The attacks fall into three
operational families. \emph{Content-tamper} attacks (\#1--\#5) mutate the
contents of an audit record --- its \texttt{probabilities} field, decision
context \texttt{ctx\_t}, selected candidate, or candidate payload --- while
leaving the leaf set of the Merkle tree intact; such attacks should be
caught by per-record commitment verification and surface as a
\texttt{commitment\_fail} signal. \emph{Leaf-removal} attacks (\#6--\#7)
remove authenticated records and should surface as
\texttt{missing\_leaves}. \emph{Synthesis/restructuring} attacks
(\#8--\#9) collapse candidate sets or inject fabricated audit records, and
should be caught by commitment verification when a leaf is altered or
added. Together the nine attacks cover silent edits,
semantics-preserving rewrites, fact supersession, knowledge-graph edge
relabeling and subgraph reanchoring, pruning, deduplication, compaction,
and poisoning. Each attack is parameterized by a strength level, and the
operation, low-level verifier signal, and literature analogue are
summarized in Table~\ref{tab:attack-defs}. The RQ4 robustness tables then
report these attacks with Rec, Mode, and $\Delta\mathrm{WK}$ to measure both
recoverable attribution and separation from the wrong-key baseline.

\input{tables/attack}

\subsection{Graphiti Backend Robustness}
\label{app:graphiti-rq4}

\input{tables/attack_signals_graphiti}

Table~\ref{tab:attack-signal-graphiti} repeats the RQ4 robustness
breakdown for the \graphiti{} backend using the same Rec, Mode, and
$\Delta\mathrm{WK}$ metrics as the \amem{} table. The Mode column keeps
the low-level diagnosis: in-place edits and synthesis-style rewrites
surface as \texttt{ComF}, while pruning and deduplication surface as
\texttt{Miss}. The $\Delta\mathrm{WK}$ columns compare each attacked
snapshot with the corresponding model's wrong-key R3 baseline, so the
table reports both recovery and key-specific attribution separation.

The graph backend shows the same qualitative pattern as \amem{} but with
backend-specific sensitivity. Content attacks have average Rec values of
$0.911$, $0.694$, and $0.489$ at strengths $0.1$, $0.3$, and $0.5$, with
average $\Delta\mathrm{WK}$ values of $+0.629$, $+0.412$, and $+0.207$.
Synthesis attacks recover $0.858$, $0.707$, and $0.583$, with
$\Delta\mathrm{WK}$ values of $+0.577$, $+0.426$, and $+0.301$.
Compaction is the most damaging synthesis case at strength $0.5$
(mean Rec $0.462$, $\Delta\mathrm{WK}$ $+0.180$), whereas poisoning remains
more recoverable because it is additive (mean Rec $0.705$,
$\Delta\mathrm{WK}$ $+0.423$).

Removal attacks preserve a recovery rate of $1.00$ over the surviving
records and retain a strong wrong-key margin ($\Delta\mathrm{WK}=+0.718$
on average), while their \texttt{Miss} mode exposes the deletion-style
lifecycle mutation. This is the desired behavior: authenticated surviving
records remain replayable, but the verifier still identifies that
evidence has disappeared from the anchored trace.

%% file: tables/metric_definitions.tex
\begin{table*}[!p]
\centering
\small
\setlength{\tabcolsep}{3.5pt}
\begin{tabular}{p{0.17\linewidth}p{0.29\linewidth}p{0.46\linewidth}}
\toprule
Axis (RQ) & Metric & Definition / Reporting Unit \\
\midrule
Utility preservation (RQ1)
  & \locomo{} F1
  & Precision and recall computed over predicted versus gold answer
    items, combined as their harmonic mean. \\
  & BLEU-1
  & Unigram precision between the generated answer and the reference
    answer, measuring lexical overlap at the token level. \\
  & $\Delta$F1 vs.\ no-watermark
  & Difference between the \memmark{} F1 and the corresponding
    \texttt{no-watermark} F1 under the same model, backend, and
    conversation. \\
\midrule
Capacity (RQ2)
  & Per-carrier entropy $H(\probdist)$
  & Shannon entropy of the self-reported candidate distribution,
    reported by carrier and averaged over the evaluated model--backend
    settings. \\
  & Per-carrier payload share
  & Fraction of embedded payload bits carried by each memory-evolution
    carrier, normalized within each model--backend setting. \\
  & bits/decision
  & Realized watermark capacity in the hyperparameter sweep: recovered
    payload bits divided by the number of watermarkable decisions. \\
\midrule
Verification (RQ3)\newline (in-record / snapshot-only)
  & R1/R2/R3 bit recovery
  & Fraction of embedded watermark bits correctly recovered from the
    available evidence under full-log R1, partial-log R2, and
    snapshot-only R3 verification. \\
  & Wrong Key
  & Observed bit-level recovery when verification is run with an
    incorrect key; for the default $K{=}4$ setting, the chance-level
    reference is $1/K{=}0.25$. \\
\midrule
Robustness (RQ4)\newline (nine memory-lifecycle attacks)
  & Rec
  & Post-attack bit recovery over the surviving verifiable records,
    reported per attack type and strength. \\
  & Mode
  & Dominant verifier signal triggered by the lifecycle attack:
    \texttt{ComF} for \texttt{commitment\_fail} and \texttt{Miss} for
    \texttt{missing\_leaves}. \\
  & $\Delta\mathrm{WK}$
  & Difference between post-attack recovery and the corresponding R3
    wrong-key baseline:
    $\mathrm{Rec}-\mathrm{WrongKey}$. Positive values indicate that the
    attacked snapshot still carries more attribution signal than an
    incorrect key. \\
\midrule
Memory integrity (RQ5)
  & Carrier Dist.
  & Counts of watermark-carrying decisions by carrier, reported as
    update:link:semantic. \\
  & Ev. Rec.
  & Fraction of QA-time gold evidence records successfully retrieved
    from memory. \\
  & Write Fail
  & Number or fraction of attempted memory writes that fail validation,
    storage, or commitment construction, as specified by the table. \\
\bottomrule
\end{tabular}
\caption{Metric definitions across the five research questions.}
\label{tab:metric-defs}
\end{table*}

%% file: sections/baseline_details.tex
\section{Details of Baselines}
\label{app:baseline-details}

This appendix clarifies the baseline definitions used in
\S\ref{sec:setup} and Table~\ref{tab:main-results}.

\begin{table*}[!p]
\centering
\scriptsize
\setlength{\tabcolsep}{4pt}
\renewcommand{\arraystretch}{1.12}
\resizebox{\textwidth}{!}{%
\begin{tabular}{p{0.13\textwidth} p{0.24\textwidth} p{0.24\textwidth} p{0.24\textwidth} c}
\toprule
Baseline & Selection rule & Evidence retained & Isolates & Backend \\
\midrule
\texttt{No-WM} &
Native backend execution; no keyed candidate selection. &
No watermark payload, reveal record, or watermark-specific Merkle leaf. &
Utility of the uninstrumented memory path. &
\makecell{\amem{}\\\graphiti{}} \\
\texttt{S.M.-Only} &
Samples from the backend distribution without embedding payload bits. &
Signed metadata, in-record sidecar, commitments, and session anchor. &
Whether authenticated metadata alone can support attribution. &
\makecell{\amem{}\\\graphiti{}} \\
\texttt{Ran.} &
Chooses randomly among the same admissible candidates, without the secret key. &
Candidate enumeration and ordinary evaluation artifacts, but no keyed payload. &
Perturbation from multi-candidate enumeration independent of watermarking. &
\makecell{\amem{}\\\graphiti{}} \\
\texttt{KGMARK} &
Applies the closest graph-watermark baseline to graph-native memory state. &
KG-level watermark evidence rather than \memmark{} state-evolution evidence. &
Comparison with a structured-memory watermark specialized to graphs. &
\makecell{\graphiti{}\\only} \\
\bottomrule
\end{tabular}%
}
\caption{\textbf{Baseline definitions}. The controls separate native
memory quality, signed provenance metadata, unkeyed candidate
randomization, and a graph-specific watermark baseline.}
\label{tab:baseline-details}
\end{table*}

\paragraph{No-WM.}
\texttt{No-WM} is the unwatermarked execution path. The memory backend,
agent harness, and evaluation pipeline are identical to the
\memmark{} runs; the only difference is that no attribution logic is
applied during memory-state evolution. It is therefore the utility
reference for RQ1 and RQ5.

\paragraph{S.M.-Only.}
\texttt{S.M.-Only} keeps the signed metadata and audit-sidecar
machinery but removes keyed selection over admissible candidates. This
control tests whether attribution can be explained by explicit metadata
alone rather than by hidden state-evolution choices. In R3, this baseline
can validate that a sidecar was signed, but it carries no keyed payload
and therefore cannot recover writer-specific state-evolution bits.

\paragraph{Ran.}
\texttt{Ran.} replaces keyed sampling with random selection among the
same admissible candidates. It therefore controls for the possibility
that any observed utility change is caused by multi-candidate
enumeration itself rather than by the secret-keyed selection rule. Since
the choice is not reproducible from $K$, it is not expected to verify as
\memmark{} evidence.

\paragraph{KGMARK.}
\texttt{KGMARK}~\cite{kgmark} is included only for \graphiti{}, where a
knowledge-graph baseline is structurally meaningful. It is omitted for
\amem{} because \amem{} does not expose graph-native edge operations of
the kind assumed by \kgmark{}. This baseline is intentionally narrower
than \memmark{}: it tests a graph-specialized watermark against a
backend-invariant state-evolution watermark.

%% file: tables/hyperparam_sweep.tex
\begin{table}[!tp]
\centering
\small
\setlength{\tabcolsep}{4pt}
\begin{tabular}{cccccc}
\toprule
$K$ & $T_{\text{enum}}$ & bits/dec. & $\Delta$F1 & R3 Recov. & Wrong Key \\
\midrule
2 & 0.7 & 0.0312 & +0.004 & 1.000 & 0.492 \\
4 & 0.5 & 0.0415 & +0.009 & 1.000 & 0.221 \\
4 & 0.7 & 0.0478 & +0.006 & 1.000 & 0.205 \\
4 & 1.0 & 0.0526 & -0.014 & 1.000 & 0.236 \\
8 & 0.7 & 0.0619 & -0.019 & 1.000 & 0.119 \\
\bottomrule
\end{tabular}
\caption{\textbf{Hyperparameter sensitivity around the default setting.}
We vary one parameter at a time around $K{=}4,T_{\text{enum}}{=}0.7$ on
Qwen3.6-flash with the \amem{} backend. Lower $K$ reduces capacity and
weakens wrong-key separation; higher $K$ increases capacity but can
perturb utility by admitting lower-quality candidates. Higher
$T_{\text{enum}}$ increases candidate diversity but also increases
semantic drift.}
\label{tab:hyperparam}
\end{table}

%% file: tables/per_conversation.tex
\begin{table*}[!t]
\centering
\tiny
\setlength{\tabcolsep}{2.4pt}
\resizebox{\textwidth}{!}{%
\begin{tabular}{c c ccc ccc cccc}
\toprule
& & \multicolumn{3}{c}{\amem{}} & \multicolumn{3}{c}{\graphiti{}} & \multicolumn{4}{c}{Watermark Metrics} \\
\cmidrule(lr){3-5}\cmidrule(lr){6-8}\cmidrule(lr){9-12}
Conv & QAs & {F1 wm} & {F1 nwm} & {$\Delta$} & {F1 wm} & {F1 nwm} & {$\Delta$} & {R1} & {R2 $r{=}0.5$} & {R3} & {Wrong Key} \\
\midrule
0 & 199 & 0.304 & 0.285 & +0.019 & 0.245 & 0.257 & -0.012 & 1.000 & 0.600 & 1.000 & 0.175 \\
1 & 105 & 0.365 & 0.393 & -0.028 & 0.303 & 0.388 & -0.084 & 1.000 & 0.475 & 1.000 & 0.212 \\
2 & 193 & 0.389 & 0.393 & -0.004 & 0.274 & 0.268 & +0.006 & 1.000 & 0.610 & 1.000 & 0.234 \\
3 & 260 & 0.270 & 0.249 & +0.021 & 0.181 & 0.257 & -0.076 & 1.000 & 0.575 & 1.000 & 0.198 \\
4 & 242 & 0.362 & 0.348 & +0.014 & 0.162 & 0.260 & -0.097 & 1.000 & 0.700 & 1.000 & 0.241 \\
5 & 158 & 0.248 & 0.283 & -0.035 & 0.250 & 0.255 & -0.005 & 1.000 & 0.500 & 1.000 & 0.194 \\
6 & 190 & 0.372 & 0.384 & -0.012 & 0.259 & 0.247 & +0.012 & 1.000 & 0.525 & 1.000 & 0.145 \\
7 & 239 & 0.331 & 0.303 & +0.027 & 0.212 & 0.216 & -0.004 & 1.000 & 0.475 & 1.000 & 0.265 \\
8 & 196 & 0.300 & 0.276 & +0.024 & 0.252 & 0.313 & -0.061 & 1.000 & 0.600 & 1.000 & 0.205 \\
9 & 204 & 0.333 & 0.300 & +0.033 & 0.211 & 0.273 & -0.061 & 1.000 & 0.500 & 1.000 & 0.179 \\
\midrule
\textbf{Mean} & 198.6 & 0.327 & 0.322 & +0.006 & 0.235 & 0.273 & -0.038 & 1.000 & 0.556 & 1.000 & 0.205 \\
\bottomrule
\end{tabular}
}
\caption{\textbf{Qwen3.6-flash per-conversation breakdown on all 10
\locomo{} samples}. F1 wm / nwm = \memmark{} vs \texttt{no-watermark}
F1 (set formula). Watermark metrics use the \amem{} backend. Wrong Key
reports observed bit-level recovery under an incorrect key; under the
default $K{=}4$ setting, the chance-level reference is
$1/K{=}0.25$. Conversation 0 is from the standalone Qwen markdown
summaries; conversations 1--9 are from the full-result JSON files.}
\label{tab:per-conversation}
\end{table*}

%% file: tables/attack.tex
\begin{table*}[!t]
\centering
\footnotesize
\renewcommand{\arraystretch}{0.94}
\setlength{\tabcolsep}{3pt}
\begin{tabular}{p{0.035\linewidth}p{0.16\linewidth}p{0.36\linewidth}p{0.17\linewidth}p{0.18\linewidth}}
\toprule
\# & Attack & Operation on the audit record & Low-level signal & Inspiration / analogue \\
\midrule
\multicolumn{5}{l}{\emph{Content-tamper attacks --- modify committed record bytes in place.}} \\
\midrule
1 & \texttt{manual\_edits}
  & Silently mutate the \texttt{probabilities} field of a record while leaving all
    other fields and the leaf set intact.
  & \texttt{commitment\_fail}
  & MemoryGraft~/ generic record tampering. \\
2 & \texttt{para.\_rewrite}
  & Append a \texttt{[PARAPHRASE]} marker to \texttt{ctx\_t}, simulating a semantics-preserving
    rewrite of the context that produced the decision.
  & \texttt{commitment\_fail}
  & RAG-WM paraphrase attack. \\
3 & \texttt{supersession}
  & Replace \texttt{selected\_candidate\_id} with a sibling candidate from the same
    carrier, simulating a newer fact overwriting an older one.
  & \texttt{commitment\_fail}
  & Graphiti native fact-invalidation chain. \\
4 & \texttt{edge\_relabel} (KG)
  & Append \texttt{[RELABEL]} to the selected candidate's \texttt{payload.text}, simulating
    relabeling of an entity--relation edge in the knowledge graph.
  & \texttt{commitment\_fail}
  & KGMark edge perturbation. \\
5 & \texttt{subgraph\_reanch.} (KG)
  & Append \texttt{[REANCHOR]} to \texttt{ctx\_t} and rotate the candidate list, simulating
    a change of root anchor for a subgraph.
  & \texttt{commitment\_fail}
  & KGMark anchor swap. \\
\midrule
\multicolumn{5}{l}{\emph{Leaf-removal attacks --- remove authenticated leaves from the trace.}} \\
\midrule
6 & \texttt{pruning}
  & Delete a fraction of audit-record leaves uniformly at random, controlled by
    the attack strength.
  & \texttt{missing\_leaves}
  & Memory-lifecycle pruning~/ KGMark subgraph removal. \\
7 & \texttt{dedup}
  & Find records duplicated by selected payload text and remove secondary
    copies, retaining only the canonical leaf.
  & \texttt{missing\_leaves}
  & Memory-lifecycle deduplication. \\
\midrule
\multicolumn{5}{l}{\emph{Synthesis/restructuring attacks --- rewrite or add records without valid openings.}} \\
\midrule
8 & \texttt{compaction}
  & Collapse the candidate set by removing one candidate, simulating multiple
    memories being merged into a single summary.
  & \texttt{commitment\_fail}
  & Memory-lifecycle compaction. \\
9 & \texttt{poisoning}
  & Inject fabricated audit records into the leaf set (additive, not deletive).
  & \texttt{commitment\_fail}
  & A-MemGuard~/ KGMark node insertion. \\
\bottomrule
\end{tabular}
\caption{The nine memory-lifecycle attacks used in RQ4. They are grouped
by lifecycle operation and list the concrete audit-record mutation, the
expected low-level verifier signal, and the motivating analogue.
Tables~\ref{tab:attack-signal-split} and~\ref{tab:attack-signal-graphiti}
then report Rec, Mode, and $\Delta\mathrm{WK}$ for these attacks.}
\label{tab:attack-defs}
\end{table*}

%% file: tables/attack_signals_graphiti.tex
\begin{table*}[!t]
\centering
\scriptsize
\setlength{\tabcolsep}{2.2pt}
\renewcommand{\arraystretch}{1.18}

\definecolor{mmGContent}{RGB}{224,242,241}
\definecolor{mmGRemoval}{RGB}{252,241,241}
\definecolor{mmGSynth}{RGB}{250,219,221}
\definecolor{gRecA}{RGB}{106,184,223}
\definecolor{gRecB}{RGB}{147,205,229}
\definecolor{gRecC}{RGB}{197,224,239}
\definecolor{gRecD}{RGB}{246,236,241}
\definecolor{gRecE}{RGB}{246,198,215}
\definecolor{gRecF}{RGB}{239,124,162}

\newcommand{\grA}[1]{\cellcolor{gRecA}\textbf{#1}}
\newcommand{\grB}[1]{\cellcolor{gRecB}\textbf{#1}}
\newcommand{\grC}[1]{\cellcolor{gRecC}\textbf{#1}}
\newcommand{\grD}[1]{\cellcolor{gRecD}\textbf{#1}}
\newcommand{\grE}[1]{\cellcolor{gRecE}\textbf{#1}}
\newcommand{\grF}[1]{\cellcolor{gRecF}\textbf{#1}}
\newcolumntype{G}{>{\centering\arraybackslash}m{2.45em}}

\resizebox{\textwidth}{!}{%
\begin{tabular}{@{}>{\centering\arraybackslash}m{4.8em}>{\centering\arraybackslash}m{5.6em}>{\centering\arraybackslash}m{4.1em}
GGG@{\hspace{2pt}}GGG@{\hspace{7pt}}
GGG@{\hspace{2pt}}GGG@{\hspace{7pt}}
GGG@{\hspace{2pt}}GGG@{}}
\toprule
& & & \multicolumn{6}{c}{\textbf{Deepseek-V4-pro}} & \multicolumn{6}{c}{\textbf{Qwen3.6-flash}} & \multicolumn{6}{c}{\textbf{GLM-5}} \\
\cmidrule(lr){4-9}\cmidrule(lr){10-15}\cmidrule(lr){16-21}
\textbf{Family} & \textbf{Attack} & \textbf{Mode}
& \multicolumn{3}{c}{\textbf{Rec}} & \multicolumn{3}{c}{\textbf{$\Delta\mathrm{WK}$}}
& \multicolumn{3}{c}{\textbf{Rec}} & \multicolumn{3}{c}{\textbf{$\Delta\mathrm{WK}$}}
& \multicolumn{3}{c}{\textbf{Rec}} & \multicolumn{3}{c}{\textbf{$\Delta\mathrm{WK}$}} \\
\midrule

\cellcolor{mmGContent}\textbf{Content} & \cellcolor{mmGContent}\textbf{Content-1} & \cellcolor{mmGContent}\texttt{ComF} & \grA{0.97} & \grC{0.65} & \grE{0.45} & \grB{+0.68} & \grD{+0.35} & \grE{+0.15} & \grB{0.82} & \grC{0.65} & \grD{0.55} & \grC{+0.47} & \grD{+0.30} & \grE{+0.20} & \grA{0.93} & \grC{0.68} & \grE{0.37} & \grA{+0.73} & \grC{+0.49} & \grE{+0.17} \\
\cellcolor{mmGContent} & \cellcolor{mmGContent}\textbf{Content-2} & \cellcolor{mmGContent}\texttt{ComF} & \grB{0.85} & \grC{0.78} & \grD{0.50} & \grB{+0.55} & \grC{+0.48} & \grE{+0.20} & \grA{0.90} & \grC{0.70} & \grD{0.50} & \grB{+0.55} & \grD{+0.35} & \grE{+0.15} & \grA{0.98} & \grC{0.76} & \grD{0.51} & \grA{+0.78} & \grB{+0.56} & \grD{+0.32} \\
\cellcolor{mmGContent} & \cellcolor{mmGContent}\textbf{Content-3} & \cellcolor{mmGContent}\texttt{ComF} & \grB{0.85} & \grD{0.60} & \grE{0.45} & \grB{+0.55} & \grD{+0.30} & \grE{+0.15} & \grA{0.90} & \grD{0.62} & \grE{0.47} & \grB{+0.55} & \grD{+0.28} & \grE{+0.12} & \grA{0.90} & \grC{0.76} & \grC{0.71} & \grA{+0.71} & \grB{+0.56} & \grC{+0.51} \\
\cellcolor{mmGContent} & \cellcolor{mmGContent}\textbf{Content-4} & \cellcolor{mmGContent}\texttt{ComF} & \grA{0.93} & \grD{0.57} & \grE{0.40} & \grB{+0.62} & \grD{+0.27} & \grE{+0.10} & \grB{0.88} & \grC{0.65} & \grE{0.42} & \grC{+0.53} & \grD{+0.30} & \grF{+0.08} & \grA{0.90} & \grC{0.68} & \grE{0.41} & \grA{+0.71} & \grC{+0.49} & \grE{+0.22} \\
\cellcolor{mmGContent} & \cellcolor{mmGContent}\textbf{Content-5} & \cellcolor{mmGContent}\texttt{ComF} & \grA{0.93} & \grC{0.68} & \grE{0.42} & \grB{+0.62} & \grD{+0.38} & \grE{+0.12} & \grA{0.95} & \grA{0.90} & \grD{0.60} & \grB{+0.60} & \grB{+0.55} & \grD{+0.25} & \grA{0.98} & \grC{0.73} & \grD{0.56} & \grA{+0.78} & \grC{+0.54} & \grD{+0.37} \\

\midrule
\cellcolor{mmGRemoval}\textbf{Removal} & \cellcolor{mmGRemoval}\textbf{Removal-1} & \cellcolor{mmGRemoval}\texttt{Miss} & \grA{1.00} & \grA{1.00} & \grA{1.00} & \grA{+0.70} & \grA{+0.70} & \grA{+0.70} & \grA{1.00} & \grA{1.00} & \grA{1.00} & \grB{+0.65} & \grB{+0.65} & \grB{+0.65} & \grA{1.00} & \grA{1.00} & \grA{1.00} & \grA{+0.80} & \grA{+0.80} & \grA{+0.80} \\
\cellcolor{mmGRemoval} & \cellcolor{mmGRemoval}\textbf{Removal-2} & \cellcolor{mmGRemoval}\texttt{Miss} & \grA{1.00} & \grA{1.00} & \grA{1.00} & \grA{+0.70} & \grA{+0.70} & \grA{+0.70} & \grA{1.00} & \grA{1.00} & \grA{1.00} & \grB{+0.65} & \grB{+0.65} & \grB{+0.65} & \grA{1.00} & \grA{1.00} & \grA{1.00} & \grA{+0.80} & \grA{+0.80} & \grA{+0.80} \\

\midrule
\cellcolor{mmGSynth}\textbf{Synthesis} & \cellcolor{mmGSynth}\textbf{Synth-1} & \cellcolor{mmGSynth}\texttt{ComF} & \grC{0.72} & \grE{0.42} & \grE{0.35} & \grC{+0.42} & \grE{+0.12} & \grF{+0.05} & \grA{0.93} & \grC{0.75} & \grE{0.45} & \grB{+0.58} & \grC{+0.40} & \grE{+0.10} & \grC{0.78} & \grC{0.68} & \grD{0.59} & \grB{+0.59} & \grC{+0.49} & \grD{+0.39} \\
\cellcolor{mmGSynth} & \cellcolor{mmGSynth}\textbf{Synth-2} & \cellcolor{mmGSynth}\texttt{ComF} & \grA{0.95} & \grA{0.93} & \grB{0.85} & \grB{+0.65} & \grB{+0.63} & \grB{+0.55} & \grA{0.93} & \grC{0.67} & \grD{0.56} & \grB{+0.58} & \grD{+0.32} & \grE{+0.21} & \grB{0.84} & \grC{0.79} & \grC{0.71} & \grB{+0.64} & \grB{+0.59} & \grC{+0.51} \\

\midrule
& & & \multicolumn{3}{c}{0.1\quad 0.3\quad 0.5} & \multicolumn{3}{c}{0.1\quad 0.3\quad 0.5}
& \multicolumn{3}{c}{0.1\quad 0.3\quad 0.5} & \multicolumn{3}{c}{0.1\quad 0.3\quad 0.5}
& \multicolumn{3}{c}{0.1\quad 0.3\quad 0.5} & \multicolumn{3}{c}{0.1\quad 0.3\quad 0.5} \\
\bottomrule
\end{tabular}}
\caption{\textbf{RQ4 -- Attack-specific recovery and wrong-key separation
on \graphiti{}.} Each row reports the attack family, a compact row alias
that maps to Table~\ref{tab:attack-defs}
(\textbf{Content-1--5} $\rightarrow$ attacks 1--5,
\textbf{Removal-1--2} $\rightarrow$ attacks 6--7, and
\textbf{Synth-1--2} $\rightarrow$ attacks 8--9), and the dominant
verifier mode
(\texttt{ComF} = \texttt{commitment\_fail};
\texttt{Miss} = \texttt{missing\_leaves}). For each model, \textbf{Rec}
is post-attack bit recovery and $\Delta\mathrm{WK}=\mathrm{Rec}-
\mathrm{WrongKey}$ compares recovery with that model's wrong-key R3
baseline at attack strengths $s\in\{0.1,0.3,0.5\}$. Positive
$\Delta\mathrm{WK}$ means the attacked snapshot retains more attribution
signal than an incorrect key.}
\label{tab:attack-signal-graphiti}
\end{table*}